\documentstyle[12pt,epsf]{article}

\begin{document}

\def\bb{b\bar{b}}
\def\bu{B^+}
\def\bd{B^0_d} 
\def\bs{B^0_s}
\def\bsbar{\overline{B^0_s}}
\def\lb{\Lambda_b}
\def\bmix{B^0 \mbox{--} \overline{B^0}}
\def\bdmix{B_d^0 \mbox{--} \overline{B_d^0}}
\def\bsmix{B_s^0 \mbox{--} \overline{B_s^0}}
\def\bsg{b\to s\,g}
\def\dmd{\Delta m_d}
\def\dms{\Delta m_s}
\def\kstar{K^{\ast 0}}
\def\kstarbar{\overline{K}^{\ast 0}}
\def\sinsqth{\sin^2\theta_W^{eff}}
\def\Zbb{Z^0 \rightarrow b\,{\overline b}}
\def\Zcc{Z^0 \rightarrow c\,{\overline c}}
\def\Zff{Z^0 \rightarrow f\,{\overline f}}
\def\Zqq{Z^0 \rightarrow q\,{\overline q}}
\def\Zuds{Z^0 \rightarrow u\,{\overline u},d\,{\overline d},s\,{\overline s}}


\pagestyle{empty}

\renewcommand{\thefootnote}{\fnsymbol{footnote}}


\begin{flushright}
{\small
SLAC--PUB--8568\\
August 2000\\}
\end{flushright}

\begin{center}
{\bf\Large Time Dependent $B_s^0 - \overline{B_s^0}$ Mixing
Using Inclusive and Semileptonic $B$ Decays
at SLD\footnote{Work supported in part by the
Department of Energy contract  DE--AC03--76SF00515.}}

\bigskip

The SLD Collaboration$^{**}$
\smallskip

Stanford Linear Accelerator Center, \\
Stanford University, Stanford, CA 94309\\
\medskip

\vspace{1.5cm}

{\bf\large
Abstract }
\end{center}
\noindent
We set a preliminary 95\% C.L. exclusion  on the oscillation frequency of
$B_s^0 - \overline{B_s^0}$ mixing using
a sample of 400,000 hadronic $Z^0$ decays collected
by the SLD experiment at the SLC between 1996 and 1998.
The analyses determine the $b$-hadron flavor at production by exploiting the
large forward-backward asymmetry of
polarized $Z^0 \rightarrow b \overline{b}$ decays
as well as information
from the hemisphere opposite that of the reconstructed $B$ decay.
In one analysis, $B$ decay vertices are reconstructed inclusively with a
topological technique,
and separation between $B^0_s$ and $\overline{B^0_s}$ decays
exploits the $B^0_s \to D_s^-$ cascade charge structure.
In the other analysis, semileptonic decays are selected and the $B$ decay
point is reconstructed by intersecting a lepton track
with the trajectory of a topologically reconstructed $D$ meson.
The two analyses are combined with a third analysis described elsewhere
to exclude the following values of the
$B_s^0 - \overline{B_s^0}$ mixing oscillation frequency:
$\Delta m_s < 7.6$ ps$^{-1}$ and
$11.8 < \Delta m_s < 14.8$ ps$^{-1}$ at the 95\% confidence level.

\vspace{2cm}

\begin{center}

{\sl Paper Contributed to the XXXth International Conference on High Energy
     Physics (ICHEP 2000), 27 Jul -- 2 Aug 2000, Osaka, Japan.}

\end{center}
\vfill

\normalsize

\pagebreak
\pagestyle{plain}

\pagebreak

\section{Introduction}

  Transitions between $B^0$ and $\overline{B^0}$ mesons
take place via second order weak interactions.
In the Standard Model,
a measurement of the oscillation frequency $\Delta m_d$ for $\bdmix$ mixing
determines, in principle, the value of the Cabibbo-Kobayashi-Maskawa
matrix element $\left| V_{td} \right|$ and constrains
the Wolfenstein parameters $\rho$
and (the CP-violating phase) $\eta$,
which are currently poorly constrained.
However, theoretical uncertainties in calculating hadronic matrix elements
are large ($\sim 20\%$~\cite{Stocchi,Hashimoto})
and thus limit the current usefulness of precise $\Delta m_d$ measurements.
Some of these uncertainties cancel when one considers the ratio
between $\Delta m_d$ and $\Delta m_s$, leading to a reduced
theoretical uncertainty ($\sim$ 4--8\%~\cite{Stocchi,Hashimoto}).
Thus, combining measurements of the oscillation frequency of
both $\bdmix$ and $\bsmix$ mixing translates into a measurement of
the ratio $|V_{td}| / |V_{ts}|$ and provides a stronger constraint
on the parameters $\rho$ and $\eta$.

  Experimentally, a measurement of the time dependence of $\bmix$
mixing requires three ingredients: (i) the $B$ decay proper time has
to be reconstructed, (ii) the $B$ flavor at production
(initial state $t = 0$) needs to be determined, as well as (iii) the $B$
flavor at decay (final state $t = t_{\rm{decay}}$).
At SLD, the time dependence of $\bsmix$ mixing has been studied using
three different methods, two of which are described below.
The third method (``D$_{\rm s}$+tracks'') is described elsewhere
(see Ref.~\cite{Dstracks}).
All methods use the same initial state flavor tag
but they use different techniques to reconstruct the $B$ decay and
tag its final state flavor.
The data consists of some 400,000 hadronic $Z^0$ decays collected with
the upgraded vertex detector (VXD3) between 1996 and 1998.
The analyses exploit the large longitudinal
polarization of the electron beam, $P_e = (73.4\pm 0.4)\%$ for 1996-98,
to enhance the initial state tag.

\section{Detector, Simulation and Event Selection}

The components of the SLD detector relevant to this analysis
are presented here.
The Liquid Argon Calorimeter (LAC) was used for triggering, event
shape measurement and electron identification.
It provides excellent solid-angle coverage
($|\cos\theta|<0.84$ and $0.82<|\cos\theta |<0.98$ in the barrel and
endcap regions, respectively).
The LAC is divided longitudinally into electromagnetic and hadronic
sections. The energy resolution for electromagnetic showers is measured
to be $\sigma/E=15\%/\sqrt{E(GeV)}$, whereas that for hadronic showers
is estimated to be $60\%/\sqrt{E(GeV)}$.
The Warm Iron Calorimeter (WIC) provides efficient muon identification
for $|\cos\theta|<0.60$.
Tracking is provided by the Central Drift Chamber (CDC)\cite{rbrb}
for charged track reconstruction and momentum measurement and the CCD pixel
Vertex Detector (VXD)\cite{vxd3}
for precise position measurements near the interaction point.
Aside from the WIC, these systems are immersed in the 0.6 T field
of the SLD solenoid.
Charged tracks reconstructed in the CDC are linked with pixel clusters in the
VXD by extrapolating each track and selecting the best set of associated
clusters\cite{rbrb}.
The track impact parameter resolutions at high momenta
are 7.8~$\mu$m and 9.7~$\mu$m in the $r\phi$ and $rz$ projections
respectively ($z$ points along the beam direction),
while multiple scattering contributions are
$33 \,\mu$m~$/(p\,{\rm sin}^{3/2}\theta)$ in both projections (where the
momentum $p$ is expressed in GeV/c).

The centroid of the micron-sized SLC Interaction Point (IP) in the $r\phi$
plane is reconstructed with a
measured precision of $\sigma_{IP} = (4 \pm 2)\, \mu$m using tracks in sets of
$\sim30$ sequential hadronic $Z^0$ decays. The median $z$ position of tracks
at their point of closest approach to the IP in the $r\phi$ plane is used to
determine the $z$ position of the $Z^0$ primary vertex on an event-by-event
basis.  A precision of $\sim20\,\mu$m on this
quantity is estimated using the $Z^0 \rightarrow b\overline{b}$
Monte Carlo (MC) simulation.

The simulated $\Zqq$ events are generated using JETSET 7.4 \cite{jetset}.
The $B$ meson decays are simulated using the CLEO $B$ decay model
\cite{CLEO-QQ} tuned to reproduce the spectra and multiplicities
of charmed hadrons, pions, kaons, protons and leptons as measured at the
$\Upsilon$(4S) by ARGUS and CLEO \cite{argcl}.
Semileptonic decays of $B$ mesons follow the ISGW model~\cite{ISGW}
including 23\% $D^{\ast\ast}$ production.
The branching fractions
of the charmed hadrons are tuned to the existing measurements
\cite{PDG96}. The lifetimes of $B$ mesons and $b$-baryons are chosen
to be $\tau_{B^+}=1.656$ ps, $\tau_{B^0}=1.562$ ps,
$\tau_{B^0_s}=1.464$ ps, and $\tau_{\Lambda_b}=1.208$ ps, according to
recent world averages \cite{LEPHF99}.
The $b$-quark fragmentation follows the Peterson {\em et al.} parameterization
\cite{Peterson}.
Finally, the SLD detector is simulated using GEANT 3.21 \cite{geant}.

 Hadronic $Z^0$ event selection requires at least 7
CDC tracks  which pass within 5~cm of the IP in $z$ at the point
of closest approach to the beam and which have
momentum transverse to the beam direction $p_\perp>$200~MeV/$c$.
The sum of the energy of the charged tracks passing these cuts
must be greater than 18~GeV.
These requirements remove background from $Z^0 \to
l^+ l^-$ events and two-photon interactions. In addition, the
thrust axis determined from energy clusters in the calorimeter must
have $\left|\cos\theta_T\right|<0.85$,
within the acceptance of the
vertex detector.
These requirements yield a sample
of $\sim 311,000$ hadronic $Z^0$ decays.

Good quality tracks used for vertex finding must have
at least two associated VXD hits and
$p_\perp >$250~MeV/$c\:$.
Additional requirements are that the tracks have either
three or more VXD hits or else satisfy the following criteria:
(i) have a CDC hit at a radius$<$50~cm,
(ii) have $\geq$23 hits to insure that
the lever arm provided by the CDC is appreciable,
(iii) extrapolate to within 1~cm of the IP in $r\phi$
and within 1.5~cm in $z$ to eliminate tracks which arise 
from interaction with the detector material,
(iv) have a $\chi^2/$d.o.f.$<8$ for both the CDC portion of the track
and the combined VXD-CDC track.

Tracks reconstructed in the vertex detector but unsuccessfully tracked
in the drift chamber are also used in the analyses. Such ``VXD-only'' tracks
are constructed from hits in all three of the vertex detector layers
and are used primarily to improve the overall vertex charge reconstruction.

  Both analyses make use of the inclusive topological vertexing
technique~\cite{zvtop} developed for $B$ lifetime~\cite{blife}
and $R_b$~\cite{rbprl} analyses to tag and reconstruct $b$-hadron decays.
The $b$ purity of the sample is increased by reconstructing
the vertex mass $M$, which includes a partial correction
for missing decay products (see Ref.~\cite{rbrb}).
This inclusive vertexing technique has been adapted for semileptonic decays
to reconstruct the $D$ decay topology (see below).

\section{Initial State Flavor Tagging}

The large forward-backward asymmetry for polarized $\Zbb$
decays is used as a tag of the initial state flavor.
The polarized forward-backward asymmetry $\tilde{A}_{FB}$ can be described by
\begin{equation}
\tilde{A}_{FB} (\cos\theta_T) = 2 A_b~{{A_e - P_e}\over{1 - A_e P_e}}
        ~{{\cos\theta_T}\over{1+\cos^2\theta_T}}~,
\label{afb}
\end{equation}
where $A_b = 0.935$ and $A_e = 0.150$ (Standard Model values),
$P_e$ is the electron beam longitudinal polarization,
and $\theta_T$ is the angle between the
thrust axis and the electron beam direction (the thrust axis is signed
such that it points in the same hemisphere as the reconstructed $B$ vertex).
Thus, left- (right-)polarized electrons tag $b$ ($\bar{b}$) quarks
in the forward hemisphere, and $\bar{b}$ ($b$) quarks
in the backward hemisphere.
Averaged over our acceptance, this yields an average correct tag
probability of 0.74 for an average electron polarization $P_e = 73\%$.
The probability for correctly tagging a $b$ quark at production
is expressed as
\begin{equation}
P_A(\cos\theta_T) = {{1 + \tilde{A}_{FB}(\cos\theta_T)}\over{2}}~.
\label{pa}
\end{equation}

 A jet charge technique is used in addition to the polarized
forward-backward asymmetry. For this tag, tracks in the hemisphere
opposite that of the reconstructed vertex are selected. These tracks
are required to have momentum transverse to the beam axis
$p_\perp > 0.15$ GeV/c, total momentum $p < 50$ GeV/c, impact parameter
in the plane perpendicular to the beam axis $\delta < 2$ cm,
distance between the primary vertex and the track at the point of
closest approach along the beam axis $\Delta z < 10$ cm, and
$|\cos\theta| < 0.90$.
With these tracks, an opposite hemisphere momentum-weighted track
charge is defined as
\begin{equation}
Q_{opp} = \sum_i q_i \left|\vec{p}_i \cdot \hat{T}\right|^\kappa~,
\label{qopp}
\end{equation}
where $q_i$ is the electric charge of track $i$, $\vec{p}_i$ its momentum
vector, $\hat{T}$ is the thrust axis direction, and $\kappa$ is
a coefficient chosen to be 0.5 to maximize the separation between $b$ and
$\overline{b}$ quarks.
The probability for correctly tagging a $b$ quark in the initial state
of the vertex hemisphere can be parameterized as
\begin{equation}
P_Q(Q_{opp}) = {{1}\over{1 + e^{\alpha Q_{opp}}}}~,
\label{pq}
\end{equation}
where the coefficient $\alpha = -0.27$, as determined using the
Monte Carlo simulation. This technique yields an average correct tag
probability of 0.66 and is independent of the polarized
forward-backward asymmetry tag.

  Finally, the tag is further enhanced by the addition of other
flavor-sensitive quantities from the hemisphere opposite that of the
selected vertex. For this purpose, the inclusive topological vertexing
technique mentioned earlier is used.
The sensitive variables are:
the total track charge and charge dipole of a topologically reconstructed
vertex,
the charge of a kaon identified in the Cherenkov Ring Imaging Detector,
and the charge of a lepton with high
transverse momentum with respect to the direction of the nearest jet.
The addition of these tags improves the average correct tag probability
by about 0.03.

\begin{figure}[htb]
  \centering
  \epsfxsize11cm
  \leavevmode
  \epsfbox{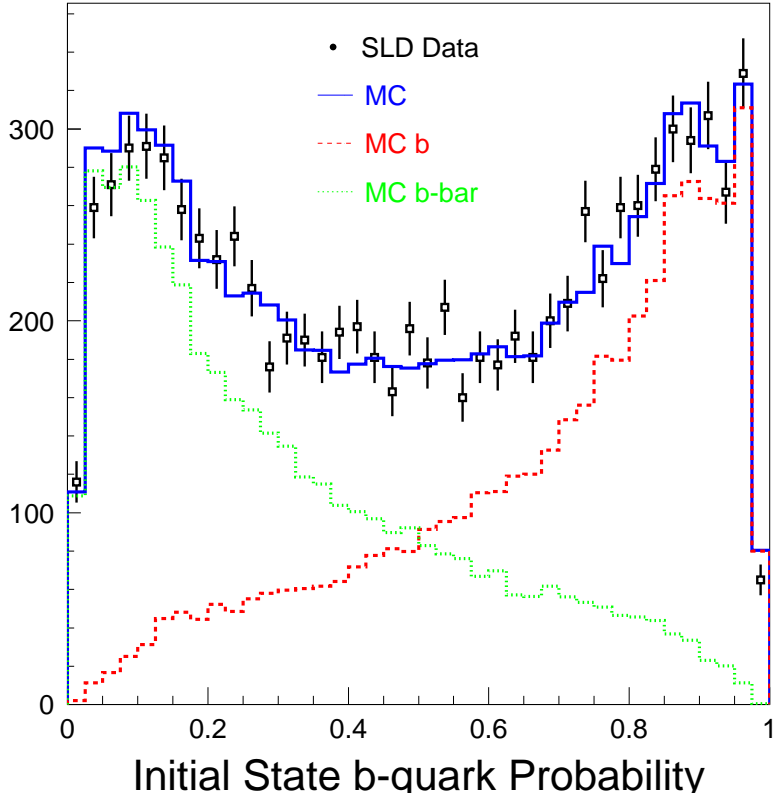}
  \caption{\it \label{fig_initag}
  \baselineskip=12pt
  Distribution of the computed initial state $b$-quark probability for
  data (points) and Monte Carlo (histograms) showing the $b$
  and $\bar{b}$ components for the events selected in the
  charge dipole analysis.}
  \baselineskip=18pt
\end{figure}

  The various tags are combined, taking correlations into account,
to form an overall initial
state tag characterized by a $b$-quark probability $P_i$.
The average correct tag probability is 0.75 for the lepton+D and 0.77
for the charge dipole analyses with 100\% efficiency.
Fig.~\ref{fig_initag} shows the $P_i$ distributions for data
and Monte Carlo in the charge dipole analysis
(see below for a description of the analysis),
and also indicates the clear separation between
$b$ and $\overline{b}$ quarks.

\section{Lepton+D Analysis}

  The lepton+``D'' analysis aims at reconstructing the $B$ and
$D$ vertex topologies of semileptonic $B$ decays.
It proceeds by first selecting event hemispheres
containing an identified lepton ($e$ or $\mu$)
with $|\cos\theta| < 0.7$.
Then, a $D$ vertex candidate is reconstructed using the inclusive
topological technique described earlier.
If multiple displaced vertices are found in the same hemisphere,
the vertex with the largest invariant mass is chosen to be the $D$ vertex
candidate.
A resultant $D$ ``track'' is created using the $D$ vertex location
and the parameters of all tracks attached to it. Furthermore,
the $D$ track error matrix is corrected to take into account the fact
that the $D$ decay is not fully reconstructed.
Finally, the $B$ decay vertex is reconstructed by intersecting the lepton
and $D$ tracks.

\begin{figure}[t]
  \vspace*{-4mm}
  \centering
  \epsfxsize12cm
  \leavevmode
  \epsfbox{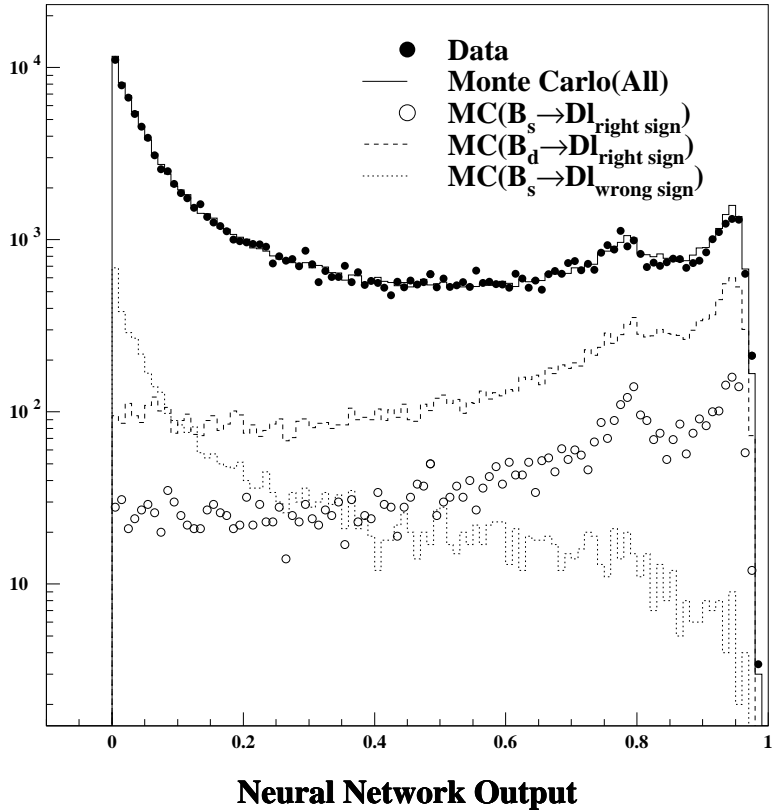}
  \caption{\it \label{fig_lepd_nnout}
  \baselineskip=12pt
  Distribution of neural network output variable
  for data (solid points) and Monte Carlo (solid histogram).
  Also shown are the Monte Carlo distributions for $\bs$ right-sign leptons
  (open circles), $\bd$ right-sign leptons (dashed histogram)
  and $\bs$ wrong-sign leptons (dotted histogram).}
  \baselineskip=18pt
\end{figure}

  A neural network is used to clean up the $D$ vertex candidates and
reduce the contamination from cascade ($b \to c \to l$)
charm semileptonic decays. The Jetnet neural network package
is used with 6 inputs and 12 hidden nodes. The inputs are
the transverse momentum of the lepton with respect to the
$B$ vertex direction (vector stretching from the IP to the $B$ vertex),
the $B$ decay length (magnitude of that vector),
the transverse momentum of the lepton with respect
to the $D$ vertex direction (vector stretching from the $B$ vertex to the $D$ vertex),
the mass $M$ of the charged tracks associated with the $B$ decay
and the distance of closest approach of the lepton to the $B$ vertex.
A distribution of the neural network
output is shown in Fig.~\ref{fig_lepd_nnout}.
A minimum cut of 0.65 is applied on the neural network output.

For this analysis, only vertices with positive
reconstructed decay length are selected.
To enhance the fraction of $\bs$ decays, the sum of
lepton $+$ $D$ vertex track charges is required to be $Q = 0$.
This enhances the
$\bs$ fraction to 16.4\% of all $b$ hadrons in the $\Zbb$ MC
(the $\bs$ production fraction in the $\Zbb$ MC is 10.0\%).
The $udsc$ contamination is 0.5\% in the final sample.

The $\bs$ fraction is further enhanced by using the Cherenkov Ring Imaging Detector
to identify kaon candidates among the $D$ vertex tracks.
For the subsample containing a lepton-kaon pair with opposite charge,
the $\bs$ fraction increases to 38.5\%
and the $udsc$ contamination remains at 0.5\%, as determined from MC.

A sample of 2087 decays is thus obtained in the 1996-98 data.
Various comparisons between data and Monte Carlo simulation were
performed which showed good agreement.
\begin{figure}[t]
  \vspace*{-4mm}
  \centering
  \epsfxsize14cm
  \leavevmode
  \epsfbox{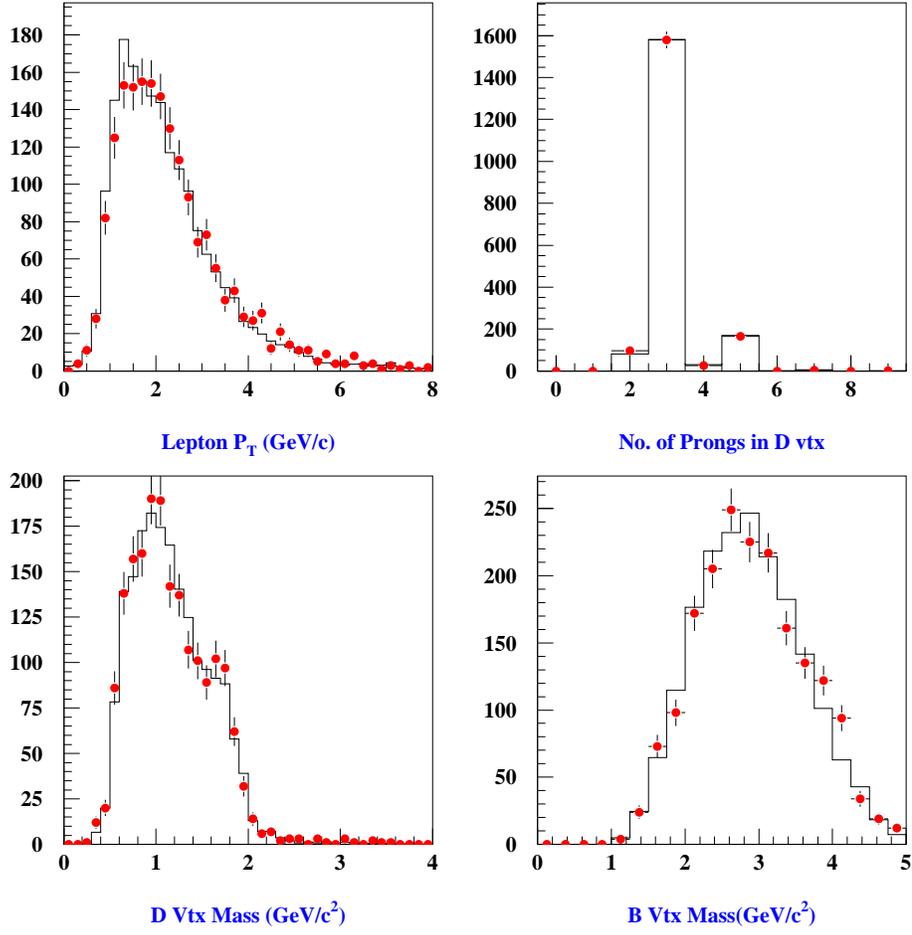}
  \caption{\it \label{fig_lepd_datamc}
  \baselineskip=12pt
  Distributions of lepton momentum transverse to the $D$ vertex trajectory,
  $D$ vertex track multiplicity, $D$ vertex mass and lepton+$D$ vertex mass
  for data (points) and Monte Carlo (histograms).}
  \baselineskip=18pt
\end{figure}
For example, Fig.~\ref{fig_lepd_datamc} shows the distributions of
lepton momentum transverse to the $D$ vertex trajectory,
$D$ vertex track multiplicity, and invariant mass of all tracks in the $D$
vertex and (assuming all tracks are pions) as well as in both $B$ and
$D$ vertices.

\begin{figure}[p]
  \hspace*{6mm}
  \centering
  \epsfxsize=11cm
  \epsfbox{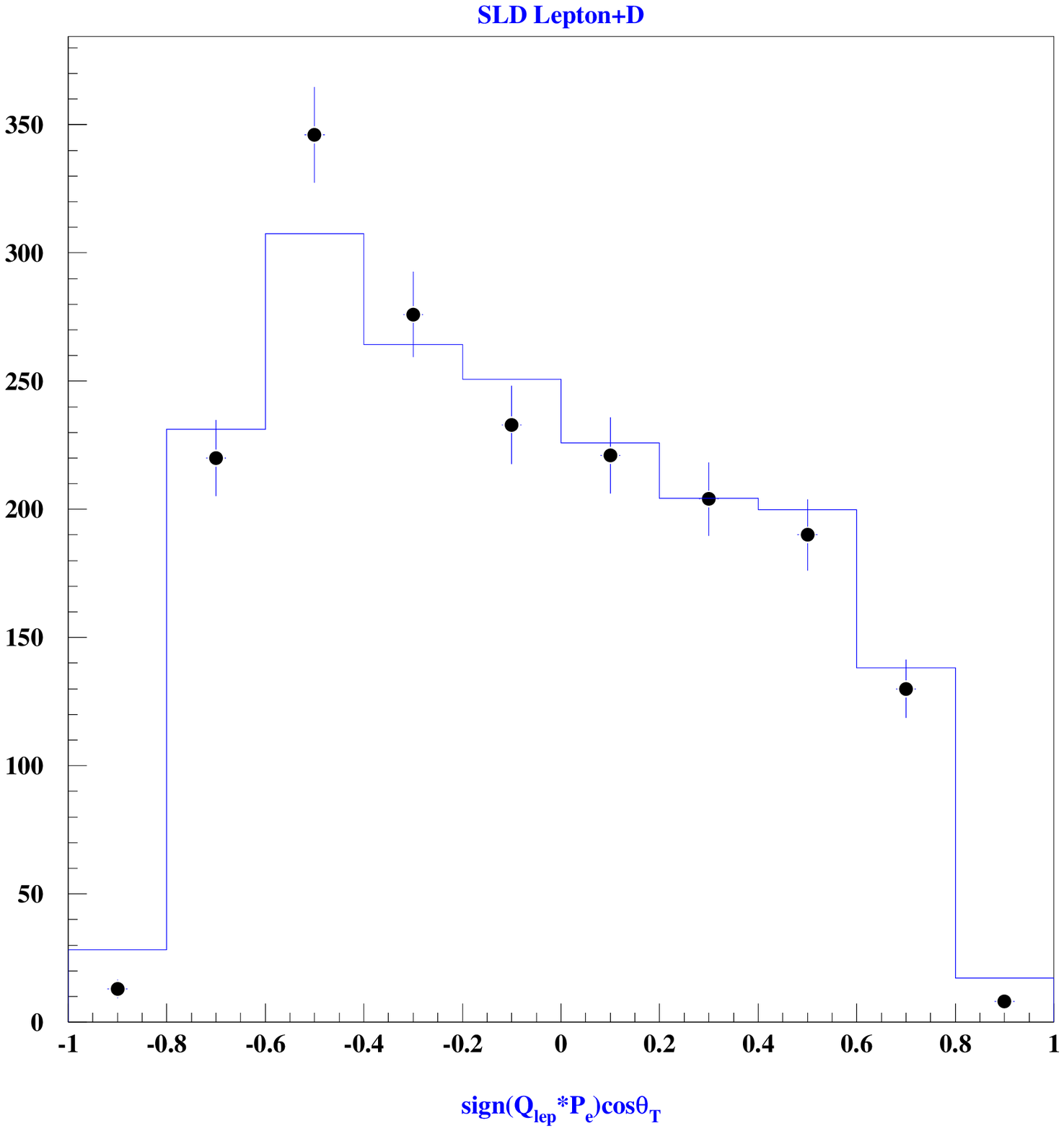}
  \caption{\it \label{fig_lepd_polasym}
  \baselineskip=12pt
  Distribution of $\cos\theta$ for the thrust axis direction
  signed by the product $(Q_{lepton} \times P_e)$
  for data (points) and Monte Carlo (histogram).}
  \baselineskip=18pt
  \hspace*{6mm}
  \centering
  \epsfxsize=10cm
  \epsfbox{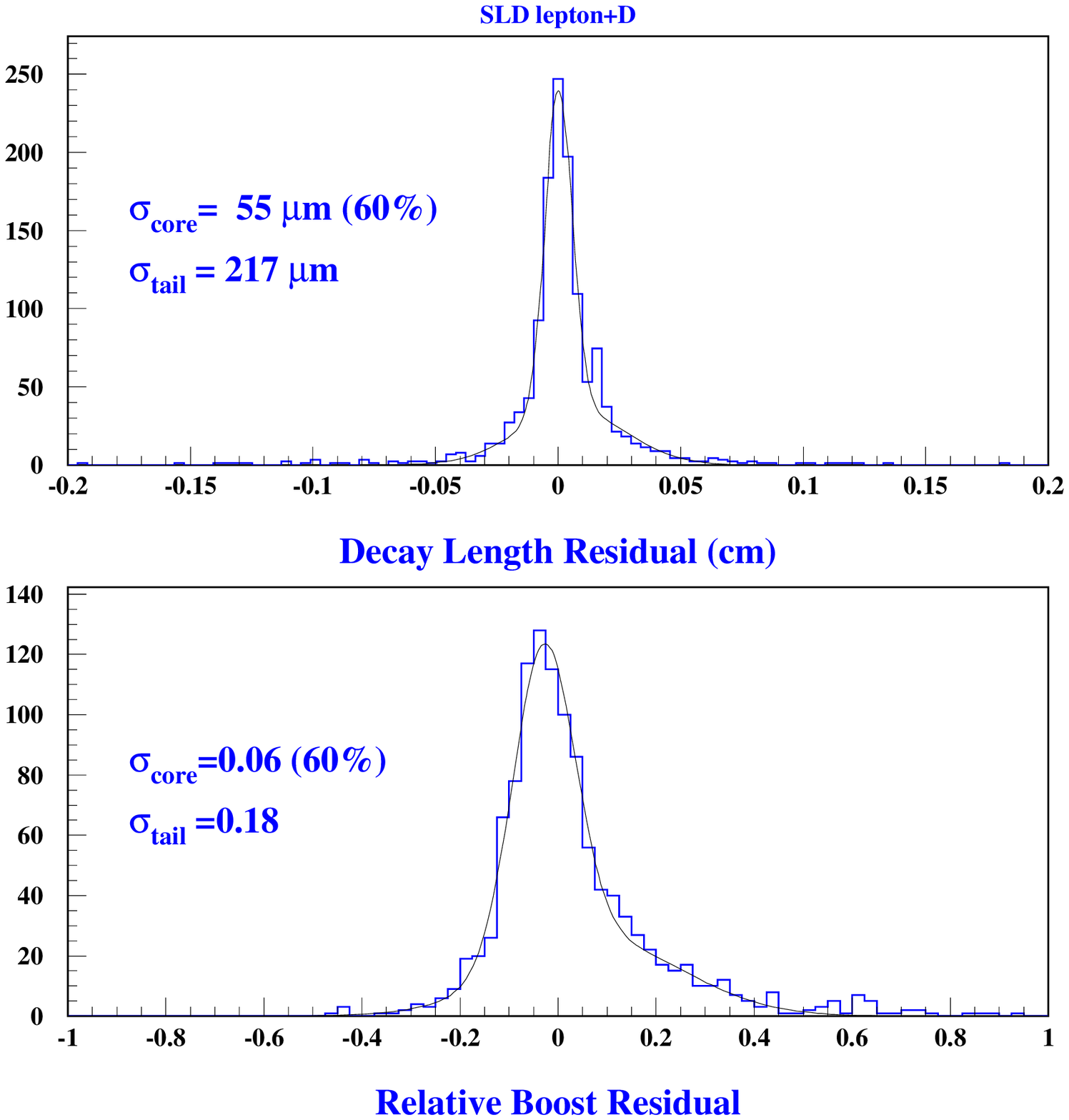}
  \caption{\it \label{fig_lepd_resol}
  \baselineskip=12pt
  Distributions of the decay length and relative boost residuals
  for $\bs$ ($b\to l$) decays in the simulation.}
  \baselineskip=18pt
\end{figure}

A powerful check of the analysis and the purity of the final state tag
is the polarization-dependent forward-backward asymmetry
shown in Fig.~\ref{fig_lepd_polasym}.
A clear asymmetry is observed, in reasonable agreement with the Monte
Carlo, indicating that the final state tag purity is adequately
modeled in the simulation.

\begin{figure}[t]
  \hspace*{6mm}
  \centering
  \epsfxsize=11cm
  \epsfbox{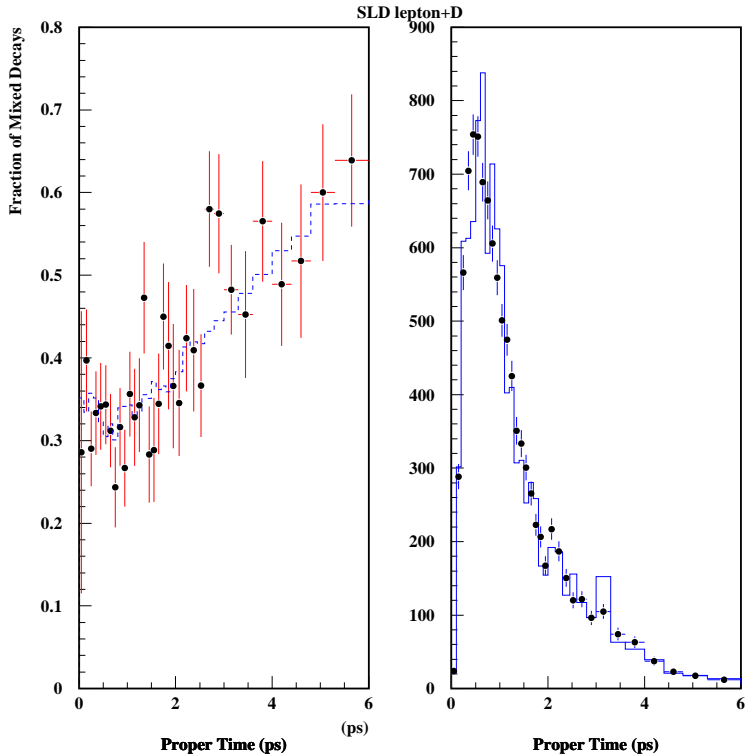}
  \caption{\it \label{fig_lepd_time}
  \baselineskip=14pt
  Distributions of the fraction of decays tagged as ``mixed'' as
  a function of reconstructed proper time (left)
  and reconstructed proper time (right) for
  the data (points) and the likelihood function (histograms).}
  \baselineskip=18pt
\end{figure}

The study of the time dependence of $\bsmix$ mixing requires
a precise determination of the $B$ decay proper time
$t = L/(\gamma\beta c)$, where $L$ is the reconstructed decay length
(distance between the IP and the $B$ vertex)
and $\gamma\beta = p_B/m_B$ is computed from the estimated $B$ momentum
$p_B$
and the known mass of the $B$ meson, $m_B$.
Reconstruction of the $b$-hadron boost uses both tracking and
calorimeter information.
A description of the reconstruction algorithm
may be found in Ref.~\cite{bfrag}.
The overall performance of the decay length and boost measurements
for $\bs$ decays proceeding via the direct $(b \to l)$ transition
is shown in Fig.~\ref{fig_lepd_resol}.
The proper time distribution is shown in Fig.~\ref{fig_lepd_time}.

  The final state $B^0$ or $\overline{B^0}$ flavor is tagged by the sign
of the lepton charge. Each decay is assigned a final state $b$-quark
probability $P_f$, defined such that $P_f > 0.5$ ($< 0.5$) corresponds
to a negatively (positively) charged
lepton which then tags the decay as $\overline{B}$ ($B$).
The magnitude of the correct tag probability depends on the sample
composition as well as on the lepton $p_T$.
The lepton sources in selected $\bs$ decays are as follows:
94.8\% $(b \to l^-)$, 2.6\% $(b \to c \to l^+)$,
1.4\% $(b \to \bar{c} \to l^-)$,
0.6\% $(b \to X^-)$ (right-sign misidentified lepton), and
0.6\% $(b \to X^+)$ (wrong-sign misidentified lepton).
The final state correct tag probability is thus 0.968.
Further enhancement of the tag is achieved by taking into account
the strong $p_T$ dependence of the various lepton source fractions.

\subsection{Likelihood Function}
\label{sec_lepd_likelihood}

  The search for the time dependence of $\bsmix$ mixing is carried out
with a likelihood analysis which includes the effect of detector smearing,
mistag of both initial and final states, selection efficiencies and
the dependence on the oscillation frequency $\dms$.
The probability that a 
meson created as a $B^0_s$ ($\overline{B^0_s}$) 
will decay as a $B^0_s$ ($\overline{B^0_s}$)  after proper time $t$ 
can be written as
\begin{equation}
P_u(t) = \frac{\Gamma}{2}\:e^{-\Gamma t}\:\left[1 +\cos(\Delta m_s\:t)\right]~,
\label{punmixed}
\end{equation}
where $\Delta m_s$ is the mass difference between the mass eigenstates,
$\Gamma$ is the average decay width of the two states 
and $P_u$ denotes the probability to remain `unmixed'.
The effects of CP violation are assumed to be small and are neglected.
Similarly, the probability that the same initial
state  will `mix' and decay as its antiparticle is
\begin{equation}
P_m(t)= \frac{\Gamma}{2}\:e^{-\Gamma t}\:\left[1 - \cos(\Delta m_s\:t)\right]~.
  \label{pmixed}
\end{equation}

  Decays are tagged as mixed or unmixed if the
product $(P_i - 0.5) \times (P_f - 0.5)$ is smaller or greater than 0,
respectively.
The probability for a decay to be in the mixed sample is expressed as:
\begin{eqnarray}
  \lefteqn{{\cal P}_{mixed}(t,\dms) = \! f_u \frac{e^{-t/\tau_u}}{\tau_u}
      \left[w^I (g_u^{dl}+g_u^{cr}+g_u^{xr})
            + (1 - w^I) (g_u^{cw}+g_u^{xw}) \right]} \nonumber \\
 & \!\! + & \!\!\frac{f_d}{2} \frac{e^{-t/\tau_d}}{\tau_d}
      \left[(g_d^{dl}+g_d^{cr}+g_d^{xr}) (1 + [2 w^I - 1] \cos\dmd t)
           + (g_d^{cw}+g_d^{xw}) (1 - [2 w^I - 1] \cos\dmd t) \right]
      \nonumber \\
 & \!\! + & \!\!\frac{f_s}{2} \frac{e^{-t/\tau_s}}{\tau_s}
      \left[(g_s^{dl}+g_s^{cr}+g_s^{xr}) (1 + [2 w^I - 1] \cos\dms t)
           + (g_s^{cw}+g_s^{xw}) (1 - [2 w^I - 1] \cos\dms t) \right]
      \nonumber \\
 & \!\! + & \!\! f_{baryon} \frac{e^{-t/\tau_{baryon}}}{\tau_{baryon}}
      \left[w^I (g_{baryon}^{dl}+g_{baryon}^{cr}+g_{baryon}^{xr})
            + (1 - w^I) (g_{baryon}^{cw}+g_{baryon}^{xw}) \right] \nonumber \\
 & \!\! + & \!\!\frac{f_{udsc}}{2}~F_{udsc}(t), \label{Pmixlepd}
\end{eqnarray}
where $f_j$ represents the fraction of each $b$-hadron type and background
($j=u,d,s$, $\Lambda$, and $udsc$ correspond to
$\bu$, $\bd$, $\bs$, $b$-baryon, and $udsc$ background),
$\tau_j$ is the lifetime for $b$ hadrons of type $j$,
$w^I$ is the initial state mistag probability,
$g_j^{dl}$, $g_j^{cr}$ and $g_j^{xr}$ are the fractions of right-sign
$(b \to l^-)$, $(b \to \bar{c} \to l^-)$ and $(b \to X^-)$ leptons,
respectively,
$g_j^{cw}$ and $g_j^{xw}$ are the fractions of wrong-sign
$(b \to c \to l^+)$ and $(b \to X^+)$ leptons,
respectively,
and $F_{udsc}(t)$ is a function describing the proper time distribution
of the $udsc$ background (a sum of two exponentials is used).
A similar expression for the probability ${\cal P}_{unmixed}$ to observe
a decay tagged as unmixed is obtained by replacing the mistag rate
$w^I$ by $1 - w^I$.

  Detector and vertex selection effects are introduced by convoluting
the above probability functions with a proper time resolution function
${\cal R}(T,t)$ and a time-dependent efficiency function $\varepsilon (t)$:
\begin{equation}
  P_{mixed}(T,\dms) = \int_{0}^{\infty} {\cal P}_{mixed}(t,\dms)~{\cal R}(T,t)
  ~\varepsilon(t)~dt~,
  \label{Pmixed}
\end{equation}
where $t$ is the ``true'' time and $T$ is the reconstructed time.
Again, a similar expression applies to the unmixed probability $P_{unmixed}$.
The resolution function is parameterized by the sum of four Gaussians:
\begin{eqnarray}
  {\cal R}(T,t)& = &
  \sum_{i=1}^{2}~\sum_{j=1}^{2}~f_{ij}~\frac{1}{\sigma_{ij}(t)\sqrt{2\pi}}
  ~e^{-\frac{1}{2}\left(\frac{T-t}{\sigma_{ij}(t)}\right)^2}~,
  \label{Resol}
\end{eqnarray}
where the index $i=1$ ($i=2$) corresponds to the core (tail) component of the
decay length resolution $\sigma_{Li}$, the index $j=1$ ($j=2$) similarly
corresponds to the core (tail) component of the relative boost resolution
$\sigma_{\gamma\beta j}/\gamma\beta$.
The various fractions are $f_{11} = 0.36$, $f_{12} = f_{21} = 0.24$,
and $f_{22} = 0.16$.
The proper time resolution $\sigma_{ij}(t)$ is a function of proper time
that depends on the measured boost $\gamma\beta$, its resolution
and the decay length resolution:
\begin{equation}
  \sigma_{ij}(t) = \left[\left(\frac{\sigma_{Li}}{\gamma\beta c}\right)^2
  + \left(t\,\frac{\sigma_{\gamma\beta j}}{\gamma\beta}\right)^2\right]^{1/2}~.
\end{equation}
For each decay, the resolution $\sigma_L$ is computed from the vertex fit
and IP position measurement errors, with a scale factor determined using
the MC simulation (the scale factor is introduced mostly to account
for the fact that the analysis does not attempt to fully reconstruct
the $D$ meson decay).
The relative boost residual $\sigma_{\gamma\beta}/{\gamma\beta}$
is parameterized as a function of the lepton $+$ $D$ vertex total
track energy, with parameters extracted from the MC simulation.
Similarly, the efficiency $\varepsilon(t)$ is parameterized using the
MC simulation.
All parameterizations are performed separately for each $b$-hadron type.
For example, the efficiency for $\bs$ decays is
given by
\begin{equation}
 \varepsilon(t) = a~\frac{1 - e^{bt}}{1 + e^{bt}} + c,
\end{equation}
with $a = 0.148$, $b = -5.7$, and $c = 0.0072$.
Furthermore, $\sigma_L$ and $\sigma_{\gamma\beta}$ resolutions
are handled separately for the main lepton sources $(b \to l)$,
$(b \to c(\bar{c}) \to l)$ and $(b \to X)$.
As a consequence, different resolution functions are used for the
different sources and the expressions for $P_{mixed}$ and
$P_{unmixed}$ are modified accordingly.

  The likelihood function is constructed from the calculated probabilities
for events tagged as mixed and unmixed.
The total likelihood for the sample is given by
\begin{equation}
  {\cal L} = \prod_{i=1}^{\# mixed} P_{mixed}
             ~\prod_{j=1}^{\# unmixed} P_{unmixed}~.
\end{equation}

\subsection{Oscillation Analysis}
\label{sec_lepd_afit}

  The study of the time dependence of $\bsmix$ mixing is carried out
using the amplitude method described in Ref.~\cite{Moser}.
Instead of fitting for $\dms$ directly, the analysis is performed
at fixed values of $\dms$ and a fit to the amplitude $A$ of the
oscillation is performed, i.e. in the expression for the mixed and unmixed
probabilities, one replaces
$\left[ 1 \pm \cos(\Delta m_s t) \right]$
with $\left[ 1 \pm A \cos(\Delta m_s t) \right]$.
This method is similar to Fourier transform analysis and
has the advantage of facilitating the combination of results
from different analyses and different experiments.

\begin{figure}[htb]
  \hspace*{6mm}
  \centering
  \epsfxsize=11cm
  \epsfbox{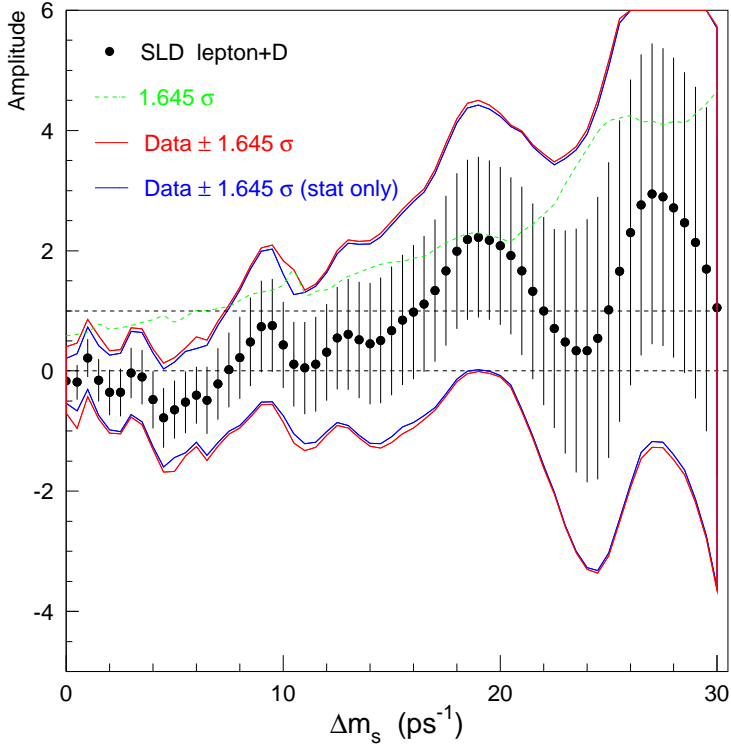}
  \caption{\it \label{fig_lepd_afit}
  \baselineskip=12pt
  Measured amplitude as a function of $\dms$ in the lepton+D analysis.}
  \baselineskip=18pt
\end{figure}
  The measured amplitude for the lepton+D analysis is shown as a function
of $\dms$ in Fig.~\ref{fig_lepd_afit}.
The measured values are consistent with $A = 0$ for the whole range of
$\dms$ up to 30 ps$^{-1}$
and no evidence is found for a preferred mixing frequency.

  Systematic uncertainties have been computed following Ref.~\cite{Moser}
and are summarized in Table~\ref{tbl_lepd_syst}
for several $\dms$ values.
\begin{table}
\caption{Measured values of the oscillation amplitude $A$ with a breakdown
   of the main systematic uncertainties for several $\dms$ values
   in the lepton+D analysis.}
\begin{center}
\begin{tabular}{lcccc}
 $\dms$                  & ~~5 ps$^{-1}$  & ~~10 ps$^{-1}$  & ~~15 ps$^{-1}$ 
                         & ~~20 ps$^{-1}$ \\
 \hline
 \vspace{0.1cm}
 Measured amplitude $A$  &      $-0.644$  &       ~0.433   &        ~0.668 
                         &       ~2.080 \\
 \vspace{0.1cm}
 $\sigma_A^{stat}$       &   $\pm 0.482$  &   $\pm 0.714$  &   $\pm 1.070$
                         &   $\pm 1.310$  \\
 \vspace{0.1cm}
 $\sigma_A^{syst}$       &
         $^{+0.113}_{-0.442}$ & $^{+0.503}_{-0.253}$ & $^{+0.249}_{-0.389}$
       & $^{+0.292}_{-0.168}$ \\
 \hline
 \vspace{0.2cm}
 $f_s = {\cal B}(\bar{b} \to \bs)$ &
         $^{-0.134}_{+0.094}$ & $^{-0.076}_{+0.296}$ & $^{-0.169}_{+0.221}$
       & $^{-0.027}_{+0.202}$ \\
 \vspace{0.1cm}
 $f_\Lambda = {\cal B}(b \to b{\rm -baryon})$ &
         $^{+0.047}_{-0.051}$ & $^{+0.065}_{-0.096}$ & $^{+0.042}_{-0.041}$
       & $^{+0.025}_{-0.027}$ \\
 \vspace{0.1cm}
 decay length resolution &
         $^{+0.003}_{-0.002}$ & $^{-0.007}_{+0.008}$ & $^{+0.028}_{-0.033}$ 
       & $^{+0.005}_{-0.044}$ \\
 \vspace{0.1cm}
 boost resolution &
         $^{-0.162}_{-0.040}$ & $^{+0.362}_{-0.113}$ & $^{+0.012}_{-0.327}$ 
       & $^{+0.206}_{-0.138}$ \\
 \vspace{0.1cm}
 $\bs$ lifetime &
         $^{+0.033}_{-0.061}$ & $^{+0.049}_{-0.176}$ & $^{+0.033}_{-0.034}$ 
       & $^{+0.002}_{-0.036}$ \\
 \vspace{0.1cm}
 $\dmd$ &
         $^{-0.093}_{-0.008}$ & $^{+0.010}_{-0.028}$ & $^{+0.000}_{-0.000}$ 
       & $^{+0.004}_{-0.005}$ \\
 \vspace{0.1cm}
 initial state tag &
         $^{+0.014}_{-0.226}$ & $^{-0.001}_{+0.003}$ & $^{-0.022}_{+0.029}$ 
       & $^{-0.025}_{+0.017}$ \\
 \vspace{0.1cm}
 ${\cal B}(b \to l)$, ${\cal B}(b \to \bar{c} \to l)$,
 ${\cal B}(b \to c \to l)$ &
         $^{+0.012}_{-0.195}$ & $^{+0.163}_{-0.000}$ & $^{+0.064}_{-0.070}$ 
       & $^{+0.019}_{-0.024}$ \\
 \vspace{0.1cm}
 lepton misidentification &
         $^{+0.008}_{-0.010}$ & $^{+0.015}_{-0.025}$ & $^{-0.005}_{+0.005}$ 
       & $^{+0.000}_{-0.004}$ \\
 \hline
\end{tabular}
\label{tbl_lepd_syst}
\end{center}
\end{table}
Uncertainties in the sample composition are estimated by
varying the fraction of $udsc$ background by $\pm 20\%$
and the production fractions of $\bs$ and $b$-baryons according to
$0.100 \pm 0.012$ and $0.099 \pm 0.017$, respectively~\cite{LEPHF99}.
Other physics modeling uncertainties are
$\tau(\bu) = 1.656 \pm 0.025$ ps,
$\tau(\bd) = 1.562 \pm 0.029$ ps,
$\tau(\bs) = 1.464 \pm 0.057$ ps,
$\tau(\lb) = 1.208 \pm 0.051$ ps,
and $\dmd = 0.480 \pm 0.020$ ps$^{-1}$.
Uncertainties in the modeling of the detector include
$\pm 10\%$ and $\pm 20\%$ variations in decay length and boost resolutions,
respectively.
Initial state tag uncertainties are estimated by varying
the correct tag probability by $\pm 0.02$ (i.e., a $\pm 10\%$ variation
of the mistag rate), corresponding to the expected contribution from
uncertainties in the measured electron beam polarization, the value of
$A_b$, and the self-calibrated jet charge analyzing power.
Final state tag uncertainties include
a $\pm 15\%$ variation in the lepton misidentification rate,
as well as the effect of uncertainties in the branching ratios
${\cal B}(b \to l) = 0.112 \pm 0.002$,
${\cal B}(b \to \bar{c} \to l) = 0.016 \pm 0.004$, and
${\cal B}(b \to c \to l) = 0.080 \pm 0.004$.
The dominant uncertainty is the $\bs$ production fraction
in $\Zbb$ events.

\section{Vertex Charge Dipole Analysis}

  The charge dipole analysis aims at selecting decays with distinct $B$
and $D$ vertices and tags the $B^0$ or $\overline{B^0}$
decay flavor based on the charge difference between them.
This analysis technique was first developed by SLD and
relies extensively on the
excellent resolution of the vertex detector.

  In the following, we first describe the algorithm used to identify primary,
secondary and tertiary vertices, then discuss details of the $\bsmix$
mixing analysis.
 
\subsection{Ghost Track Algorithm}

    The $B$ decay flavor tag with the charge dipole relies on the
  kinematic fact that the boost of the $B$ decay system carries the
  cascade charm decay downstream from the $B$ decay vertex. Monte Carlo
  studies show that in $B$ decays producing a single $D$ meson
  the cascade $D$ decays on average $4200\:\mu$m
  from the IP, while the intermediate $B$ vertex is displaced on average only
  $46\:\mu$m transversely from the line joining the IP to the $D$ decay
  vertex. This kinematic stretching of the $B$ decay chain into an
  approximately straight line is exploited by the ghost track algorithm.
  This new algorithm has two stages and operates on a given set of
  selected tracks in a jet or hemisphere. First, the best
  estimate of the straight  line from the IP
  directed along the $B$ decay chain is found.  
  This line is promoted to the status of a track by assigning it a
  finite width. This new track, regarded as
  the resurrected image of the deceased $B$ hadron, is called the ``ghost''
  track. Secondly, the selected tracks are vertexed with the ghost track
  and the IP to build up the decay chain along the ghost direction. Both
  stages are now described in more detail.

    Given a set of tracks in a hadronic jet or hemisphere
  a new track G is created 
  with the properties that it is a straight line from the IP directed
  along the jet or thrust axis
  and has a constant resolution width of $25\:\mu$m in
  both $r\phi$ and $rz$.
  For each track $i$ a vertex is formed with 
  track G and the vertex location ${\bf r_i}$, fit
  $\chi^2_i$ and L$_i$ are determined
  (L$_i$ is the longitudinal displacement from the IP  of
  ${\bf r_i}$ projected onto the direction of track G). This is calculated for
  each of the tracks and the summed $\chi^2$ is formed:
  \begin{equation}
   \chi^2_S = \sum_i \begin{array}{ll}
                     \chi^2_i        & \;\;\; \mbox{L}_i \geq 0.0    \\
                 (2{\chi^2_0}_i - \chi^2_i)  & \;\;\; \mbox{L}_i < 0.0
                     \end{array}
    \label{min1}
  \end{equation}
   where ${\chi^2_0}_i$ is the $\chi^2_i$ of track $i$ to track G determined
   at L$_i = 0$ rather than at the best fit vertex location. The aim is to
   construct this quantity, $\chi^2_S$, such that when the direction of G is
   varied the minimum of $\chi^2_S$ provides the best
   estimate of the $B$ decay direction. If the initial direction
   is a relatively
   long way from the $B$ line of flight, some or all of the decay tracks
   may vertex with G with a negative value of L$_i$.
   In this case the $2{\chi^2_0}_i - \chi^2_i$ term above
   helps to push track G towards the $B$ flight path as $\chi^2_S$ is
   minimized. This first minimization using equation~\ref{min1}
   is designed for this purpose. (Note
   that the contribution of each track as $\chi^2_S$ is minimized changes
   in a continuous manner even if L$_i$ changes sign since 
   $\chi^2_i = {\chi^2_0}_i$ at L$_i = 0$.)
                   
      The value of $\chi^2_S$ is recalculated as track G is rotated
   (about the pivot at the IP) 
   incrementally in ever decreasing angular steps $\delta \theta$ and
   $\delta \phi$ until the minimum is found within the required 
   precision ($<0.1\:$mrad, i.e. within 1$\:\mu$m at 1 cm from the IP).
   The width of track G is set such that the maximum $\chi^2_i = 1.0$
   for all tracks with L$_i > 0$
   (if this is less than 25$\mu$m, it is restored to 25$\mu$m).
   The track G is  now consistent with all
   potential $B$ decay candidate tracks (L$_i > 0$) at the level
   $\chi^2_i \leq 1.0$. In other words, the new width of G measures the
   degree to which the tracks conform to a straight line decay chain.
   A second iteration in $\delta \theta$,$\delta \phi$ now takes place
   with the summed $\chi^2$ redefined as:
  \begin{equation}
   \chi^2_S = \sum_i \begin{array}{ll}
                     \chi^2_i    & \;\;\; \mbox{L}_i \geq 0.0    \\
                    {\chi^2_0}_i & \;\;\; \mbox{L}_i < 0.0
                     \end{array}
    \label{min2}
  \end{equation}
    which is not sensitive to any spurious background track with a negative
    value of L$_i$ which might otherwise perturb the direction of track G.
    After finding the new minimum of $\chi^2_S$ the width of G is again
    recalculated such that $\chi^2_i \leq 1.0$ for all tracks $i$ with
    L$_i > 0$. Again this width is required to be at least 25$\:\mu$m.
    Track G is now directed along the best guess of the $B$ decay line of
    flight and has a width such that it is consistent with potential
    $B$ decay tracks in the jet, track G is now called the ``ghost'' track.

      The second stage of the algorithm begins by defining a fit probability
    for a set of tracks to form a vertex with each other and with the
    ghost track (or IP). This probability then measures the likelihood 
    of the set of tracks both belonging to a common vertex {\em and}
    being consistent with the ghost track (or IP) and hence forming a
    part of the $B$ decay chain. These probabilities are determined from
    the fit $\chi^2$ which is in turn determined algebraically from the
    parameters of the selected tracks and the ghost track
    (or the $7\times 7\times 30 \mu$m$^3$ ellipsoid assumed for the IP).
    The earlier requirement that each
    L$_i > 0$ track makes a $\chi^2_i \leq 1.0$ with the ghost track
    has the effect that the fit probabilities have the desired property
    of having an approximately flat distribution from 0.0 to 1.0 for
    genuine vertices, independent of both multiplicity and decay length.
    This property also relies on the choice of the number of degrees
    of freedom as 2N$-2$ (or 2N) when fitting N tracks together
    with the ghost track (or IP).
    Fake vertices peak at probability close to 0.0. 

     For a set of N tracks, there are initially N+1 candidate
    vertices (N 1-prong secondary vertices and a bare IP).
    A matrix of track $i$ -- track $j$ associations is
    constructed to store the calculated probabilities of each
    candidate vertex pair fitted together with the ghost track. A further 
    column and row is added to the matrix to store the probabilities of
    each track fit with the IP ellipsoid. The upper triangle 
    of the matrix (i.e. the $ij$ ($i<j$) elements) stores the
    probabilities while the lower triangle (initialized with $ij$
    ($i>j$) elements set to 0.0) indelibly records which tracks 
    (and IP) have been assigned together in a common vertices as
    the algorithm progresses.
      Once the upper triangle has been filled, the highest
    probability in the matrix table is found and the corresponding
    candidate vertex pair are from then on tied together
    in a new candidate vertex for all future computations
    by flagging the corresponding lower triangle elements
    of the matrix with non-zero values. The upper triangle of
    the matrix is now
    refilled taking into account the associations that
    have so far been made, the new maximum probability is 
    found, and the corresponding subset of the tracks and IP is tied
    together. At each 
    iteration of combining the maximum probability matrix element
    contributors, the number of candidate vertices decreases by one.
    The iterations continue until the maximum probability is less than
    1\%.  At this point the tracks and IP have been divided into
    unique subsets by the associations thereby defining topological
    vertices.

      Jets or hemispheres in which three vertices are found --
    the primary (which includes
    by definition the IP), a secondary and a tertiary -- are used for the
    charge dipole analysis. The secondary vertex is identified as the $B$
    decay vertex and the tertiary as the cascade charm decay. As well as
    improving the purity and efficiency of the dipole reconstruction
    (by requiring the vertices be consistent with a single line of flight)
    the ghost track algorithm has the additional advantage of allowing
    the direct reconstruction of 1-prong vertices, including the
    topology consisting of 1-prong B and D decays.

Finally, VXD-only tracks are attached to the $B$ decay chain
to improve the overall charge reconstruction but
they are not used in the determination of the
$B$ and $D$ vertex positions.
The attachement criteria rely on variables $\tilde{T_i}$ and
$\tilde{L_i}$ for each track $i$, as defined below.
A vertex axis is formed by a straight line joining the IP
to a vertex combining both secondary and tertiary tracks.
The 3-D distance of closest approach of the track to the vertex axis,
$\tilde{T_i}$,
and the distance from the IP along the vertex
axis to this point, $\tilde{L_i}$, are calculated.
Track $i$ is attached to the vertex if $\tilde{T_i} < 0.1$~cm,
$\tilde{L_i} > 0.025$~cm and $0.25 < \tilde{L_i}/\tilde{L} < 2$
(where $\tilde{L}$ is the distance
between the IP and the combined vertex).
An average of 0.2 VXD-only tracks is added per decay, to be compared with
an average of 5.0 VXD+CDC tracks per decay (for $M > 2$ GeV/c$^2$).
The VXD-only tracks attached to the $B$ decay chain are further
attached to either the $B$ or $D$ vertex according to their longitudinal
displacement $\tilde{L_i}$: tracks with $\tilde{L_i} < L_B + 0.5(L_D - L_B)$
are attached to the $B$ vertex and all others are attached to the $D$ vertex,
$L_B$ ($L_D$) is the distance between the IP and the $B$ ($D$) vertex.

\subsection{Event Selection}

Hemispheres containing both a secondary and a tertiary vertex are selected for
the charge dipole analysis.
Furthermore, the invariant mass computed using all secondary and
tertiary vertex tracks is required to be $M > 2$ GeV/c$^2$ (the computed
mass includes a partial correction for missing decay products)
and the total track charge $Q_{tot}$
(from both secondary and tertiary vertices)
is required to be zero to enhance the fraction of $\bs$ decays in the sample
and to increase the
quality of the charge difference reconstruction for neutral $B$ decays.
As mentioned in the previous section, the (secondary) vertex that is closer
to the IP is labelled ``$B$'' and that further away (tertiary) is labelled
``$D$.''
A ``charge dipole'' is defined as
$\delta Q \equiv D_{BD} \times SIGN (Q_D - Q_B)$,
where $D_{BD}$ is the distance between the two vertices and
$Q_B$ ($Q_D$) is the charge of the $B$ ($D$) vertex.
Positive (negative) values of $\delta Q$ tag $\overline{B^0}$ ($B^0$)
decays.
Requirements on the vertices are:
$250\:\mu$m $< D_{BD} < 1$ cm, $D$ vertex mass $< 2.0$ GeV/c$^2$
(assuming all tracks are pions),
$B$ vertex decay length $L > 0$, $Q_B \neq Q_D$,
``ghost'' track width $< 300\:\mu$m and
cosine between the straight line connecting the IP and the $B$ vertex, and
the nearest jet axis direction $< 0.9$.
The decay is rejected if any attached VXD-only track
has $p_\perp > 4$ GeV/c,
since in that case the charge is not reliably reconstructed.
MC studies indicate that, after these selection cuts,
the track assignment to the $B$ ($D$) vertex
is 84\% (86\%) correct for $\bs$ decays containing one $D$ meson
in the final state,
i.e. 84\% (86\%) of all tracks in the $B$ ($D$) vertex originate from
the decay point of the $B$ ($D$) meson.
For all data and MC events, hemispheres already containing a vertex
selected by the lepton+D or D$_{\rm s}$+track analyses are removed to keep
the analyses statistically uncorrelated.
The $udsc$ background is further suppressed by demanding
that the event contains either an opposite hemisphere topological vertex
with $M > 2$ GeV/c$^2$ or at least 3 tracks with positive 2-D impact
parameter $> 3\:\sigma$.
The $udsc$ fraction is thus reduced to 2.2\%.

\begin{figure}[p]
  \hspace*{6mm}
  \centering
  \epsfxsize=12cm
  \epsfbox{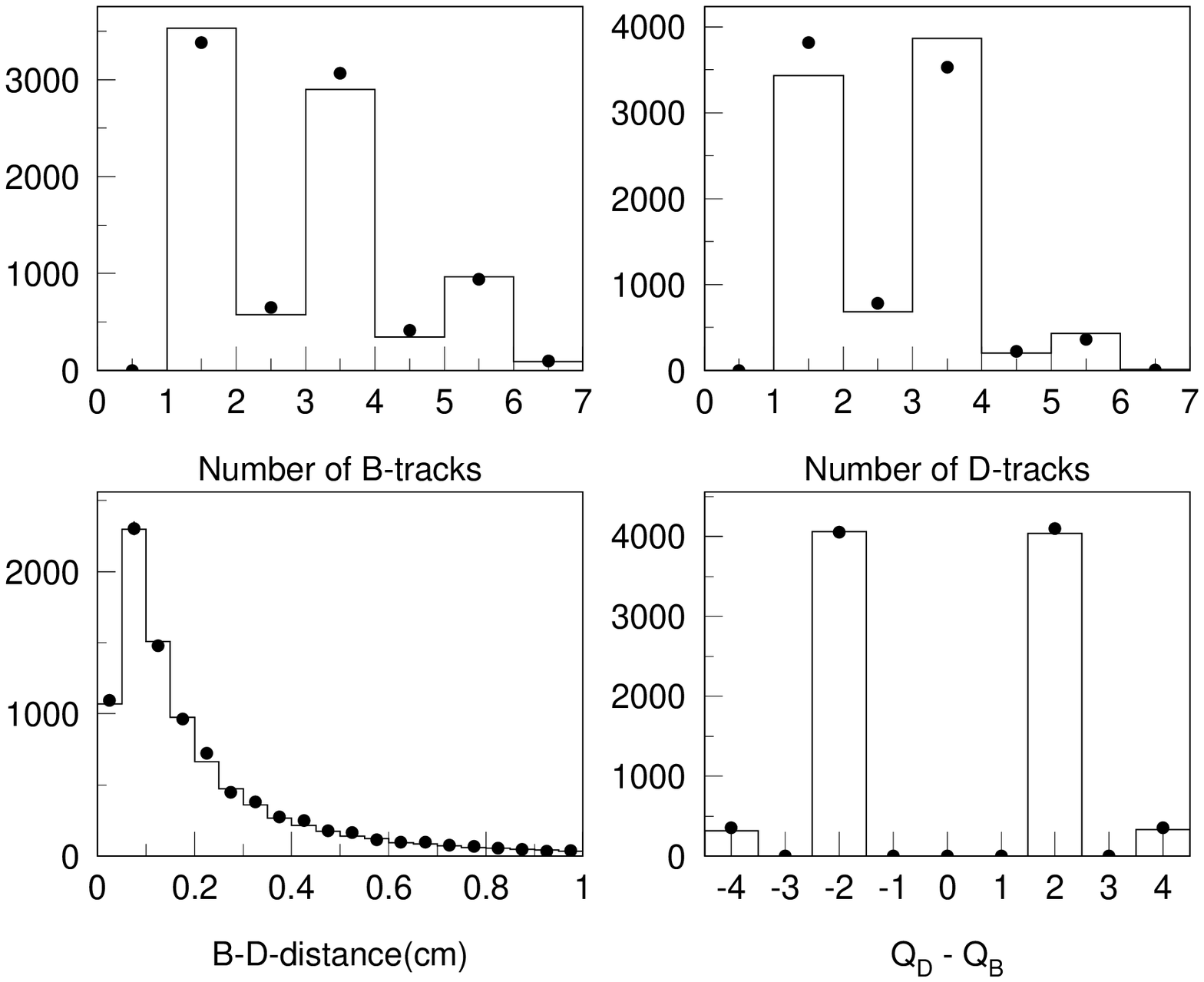}
  \caption{\it \label{fig_dipo_datamc}
  \baselineskip=12pt
  Distributions of $B$ and $D$ vertex track multiplicity, as well as
  distance and charge difference between $B$ and $D$ vertices for
  data (points) and Monte Carlo (histograms)
  in the charge dipole analysis.}
  \baselineskip=18pt
  \vspace*{1mm}
  \hspace*{6mm}
  \epsfxsize=10cm
  \epsfbox{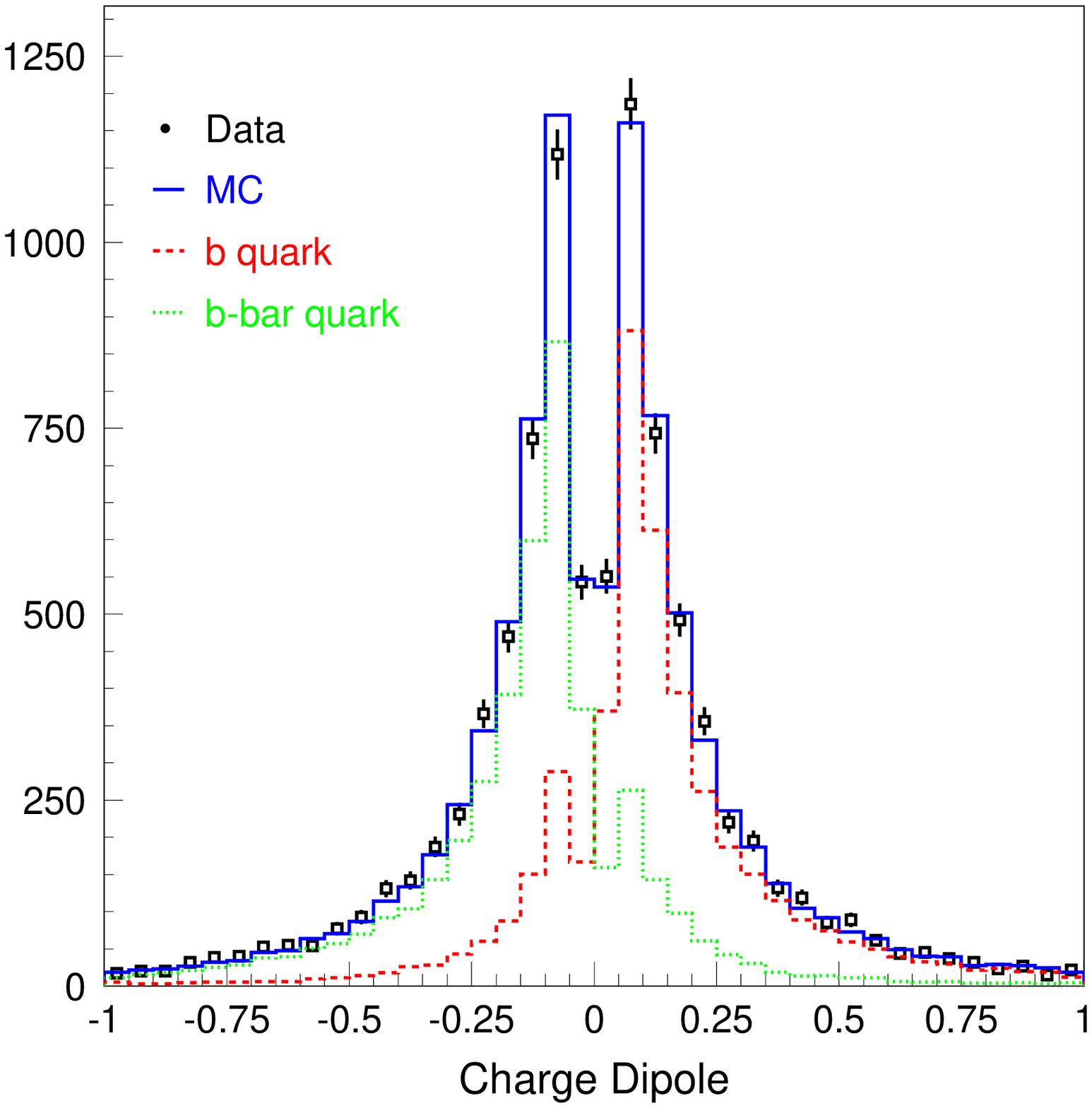}
  \caption{\it \label{fig_dipo_distr}
  \baselineskip=12pt
  Distribution of the charge dipole for
  data (points) and Monte Carlo (solid histogram).
  Also shown are the contributions from $b$ hadrons containing
  a $b$ quark (dotted histogram) or a $\bar{b}$ quark (dashed histogram).}
  \baselineskip=18pt
\end{figure}

  Applying all the above cuts, a sample of 8556 decays is selected
in the 1996-98 data.
Figure~\ref{fig_dipo_datamc} shows distributions of the
$B$ and $D$ vertex track multiplicities, as well as the
distance and charge difference between $B$ and $D$ vertices in the
selected sample.
Good agreement between data and MC is obtained.
A slight discrepancy in the $D$ vertex track multiplicity is apparent
but was determined to have negligible impact on the analysis.
Requiring that the total track charge be zero boosts the
$\bs$ fraction from its assumed production value of 10.0\% to 15.3\%.
Figure~\ref{fig_dipo_distr} displays the distribution of
charge dipole $\delta Q$
for the data sample and also indicates the separation between
$b$ hadrons containing $b$ or $\bar{b}$ quarks in the MC.

\begin{figure}[p]
  \hspace*{6mm}
  \centering
  \epsfxsize=10cm
  \epsfbox{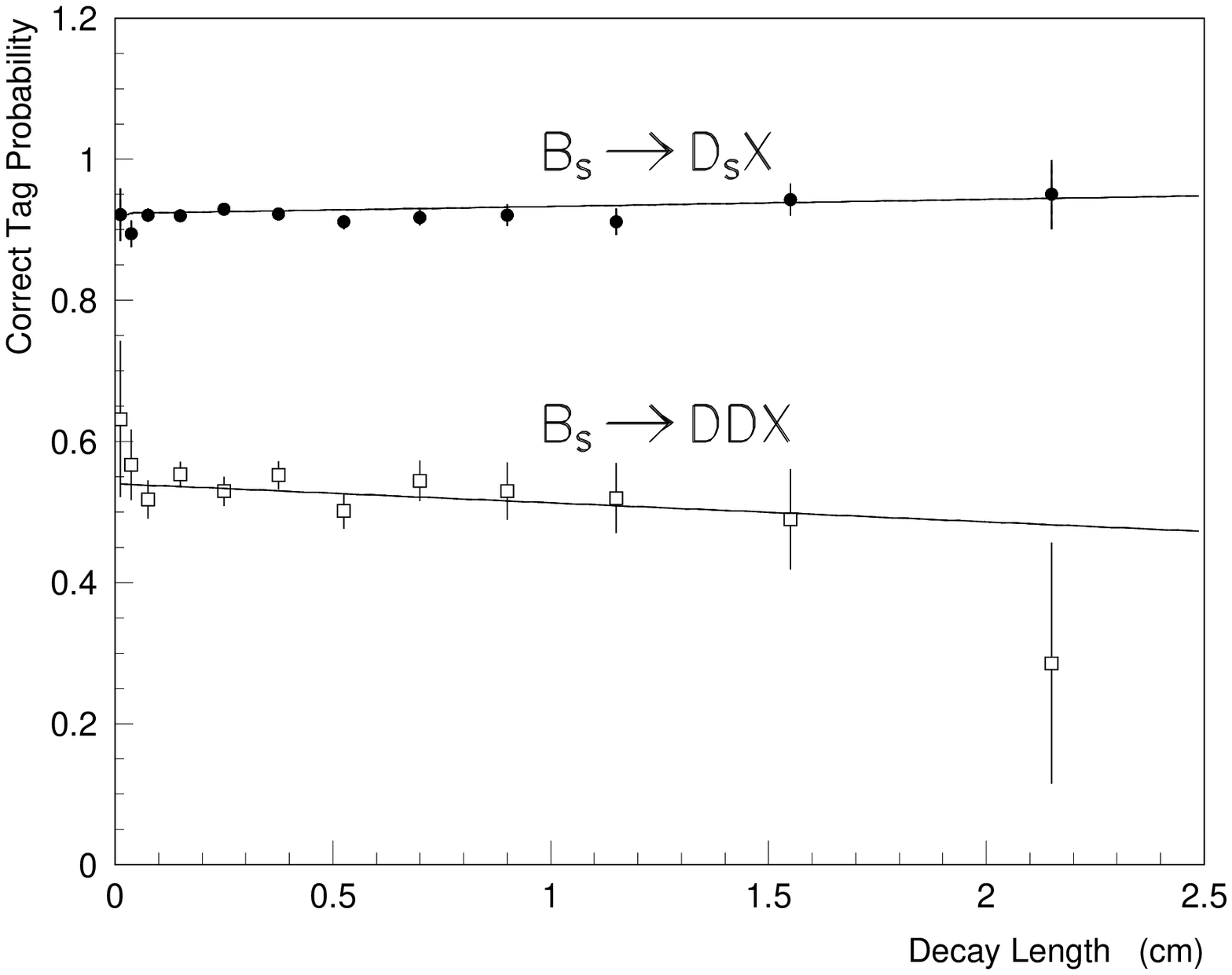}
  \caption{\it \label{fig_dipo_tagpur}
  \baselineskip=12pt
  Charge dipole correct tag probability as a function of reconstructed
  decay length in simulated $\bs \to D_s X$ (solid circles) and
  $\bs \to D \overline{D} X$ (open squares) decays.
  The functions are the fit results used to parametrize the charge dipole
  correct tag probability as a function of decay length.}
  \baselineskip=18pt
  \vspace*{4mm}
  \hspace*{6mm}
  \centering
  \epsfxsize=12cm
  \epsfbox{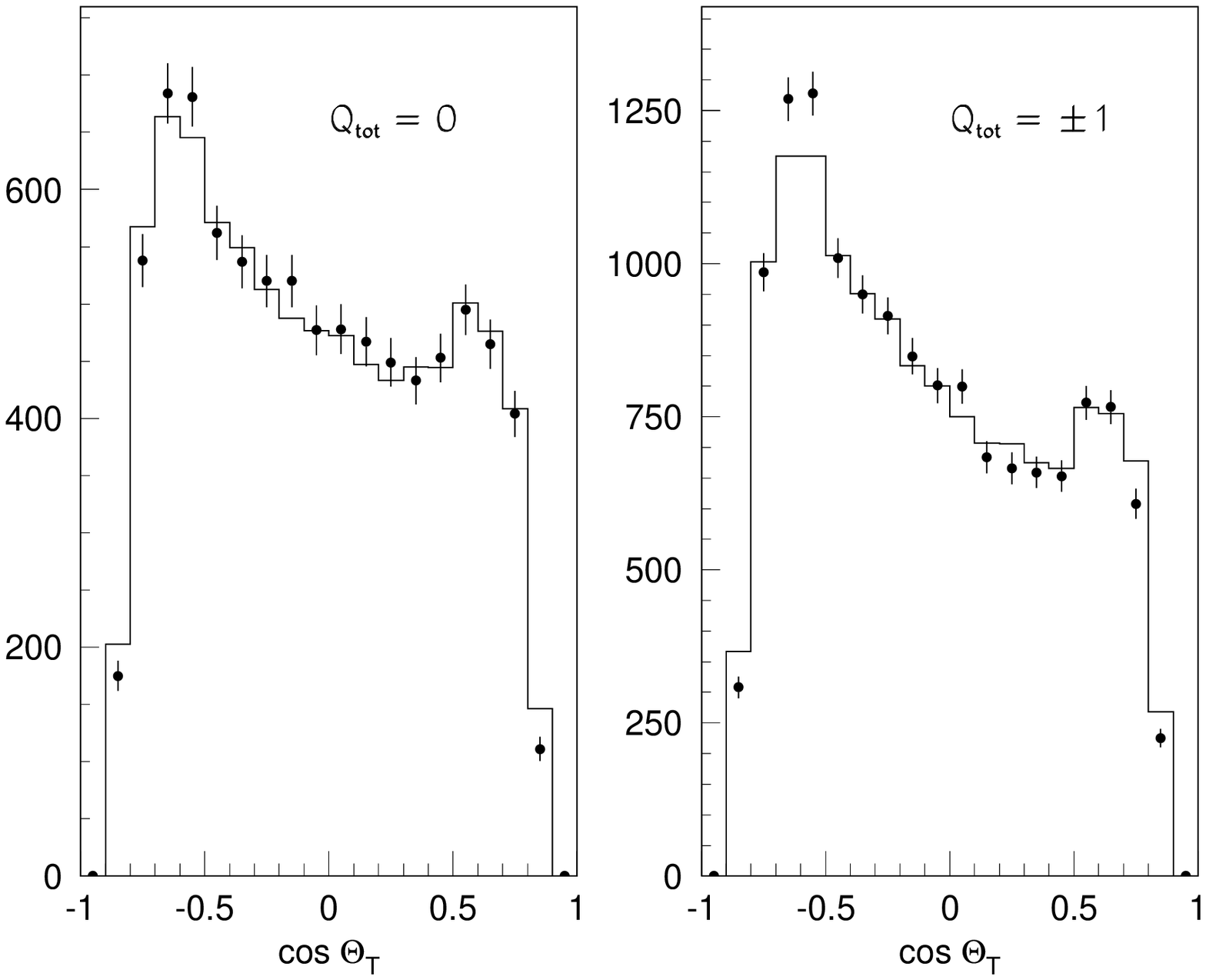}
  \caption{\it \label{fig_dipo_polasym}
  \baselineskip=12pt
  Distributions of $\cos\theta$ for the thrust axis direction
  signed by the product $(\delta Q \times P_e)$
  for data (points) and Monte Carlo (histograms)
  in subsamples with $Q_{tot} = 0$ and $Q_{tot} = \pm 1$ for
  the charge dipole analysis.}
  \baselineskip=18pt
\end{figure}
The average correct tag probability for the charge dipole tag is 0.76
for selected $\bs$ decays
and is parameterized as a function of decay length, as shown in
Fig.~\ref{fig_dipo_tagpur}.
Furthermore, the correct tag probability depends on the charm content
of the decay products. Thus, the correct tag probability is parameterized
separately for $\bd$ and $\bs$ decays into five different final states:
$D^0 X$, $D^+ X$, $D_s X$, charmed hadron $X$, and
$D \overline{D} X$ (this last category also incorporates charmonium
production, i.e. it includes all $b \rightarrow c \overline{c} s$ decays).
For example, the correct tag probability is 0.88 for $\bs \to D_s X$ decays
but only 0.53 for $\bs \to D \overline{D} X$ decays.
The above correct tag probabilities are extracted from the MC simulation
and include a multiplicative dilution factor $S_D = 0.95$.
This factor is conservatively assigned to reproduce the overall
fraction of decays tagged as mixed observed in the data, as well as to
properly describe the time evolution of $\bdmix$ mixing in the data
(see below).

\begin{figure}[thb]
  \vspace*{4mm}
  \hspace*{6mm}
  \centering
  \epsfxsize=12cm
  \epsfbox{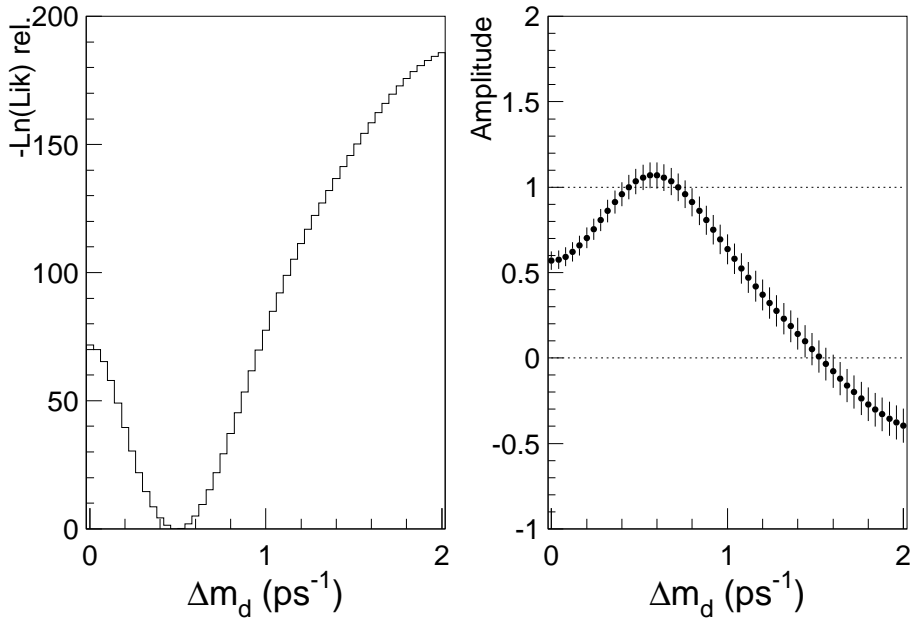}
  \caption{\it \label{fig_dipo_bdmix}
  \baselineskip=12pt
  Log likelihood as a function of $\dmd$ and measured amplitude
  as a function of $\dmd$ for the charge dipole analysis.}
  \baselineskip=18pt
\end{figure}

As hadronic decays of $B$ mesons are not as well known as semileptonic
decays, it is important to check the correct tag probability estimated using
measured quantities like the polarization-dependent forward-backward
asymmetry shown in Fig.~\ref{fig_dipo_polasym}.
Good agreement with the MC is observed, indicating that the
charge dipole correct tag probability is well modeled.
It should be noted that this asymmetry is diluted by both
initial and final state mistags and by $\bmix$ mixing.
The dilution due to mixing can be reduced by selecting
vertices with total charge $Q_{tot} = \pm 1$, in which case
a stronger asymmetry is observed (see Fig.~\ref{fig_dipo_polasym}).
Another useful test of the charge dipole tag in $\bd$ decays is the
measurement of the time dependence of $\bdmix$ mixing.
This has been checked using the full likelihood analysis described in
the following section. Fitting for the $\bdmix$ mixing frequency
yields $\dmd = 0.495 \pm 0.032$ ps$^{-1}$ (statistical error only),
see Fig.~\ref{fig_dipo_bdmix}.
This value agrees well with the latest world
average value of $0.485 \pm 0.015$ ps$^{-1}$~\cite{golutvin}.
\begin{figure}[thb]
  \hspace*{6mm}
  \centering
  \epsfxsize=9cm
  \epsfbox{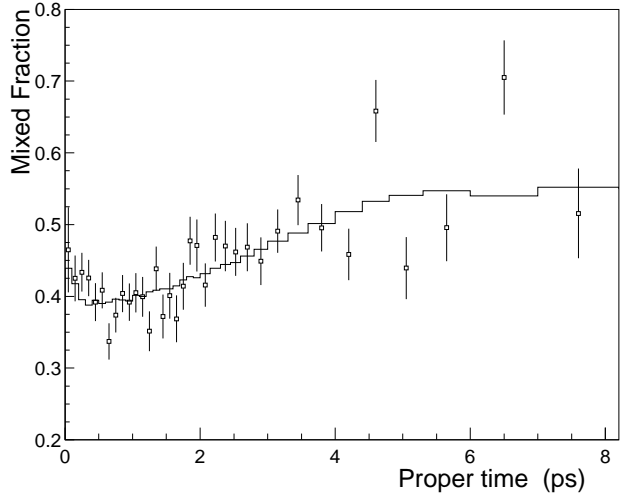}
  \caption{\it \label{fig_dipo_time}
  \baselineskip=12pt
  Distributions of the fraction of decays tagged as ``mixed''
  for the data (points) and the likelihood function (histograms)
  in the charge dipole analysis.}
  \baselineskip=18pt
\end{figure}
The mixed fraction as a function of proper time is displayed
in Fig.~\ref{fig_dipo_time}.
The figure shows the expected increase in the fraction of decays
tagged as mixed for decays at small proper time.
Most of this effect originates from misreconstructed $\bu$ decays
near the IP which tend to have random final state tags.

\subsection{Likelihood Function}

  The $\bsmix$ mixing fit for the charge dipole analysis is performed
in a way similar to the lepton+D analysis.
Decays are tagged as mixed or unmixed if the
product $(P_i - 0.5) \times (P_f - 0.5)$ is smaller or greater than 0,
respectively.
The probability for a decay to be in the mixed sample is expressed as:
\begin{eqnarray}
  \lefteqn{{\cal P}_{mixed}(t,\dms) = \! f_u \frac{e^{-t/\tau_u}}{\tau_u}
      \left[w^I (1 - w^F_u) + (1 - w^I) w^F_u \right]} \nonumber \\
 & \! + & \!\!\frac{f_d}{2} \frac{e^{-t/\tau_d}}{\tau_d}
      \left(\sum_{k=1}^{5}~g_{dk}
      \left[(1-w^F_{dk}) (1 + [2 w^I - 1] \cos\dmd t)
           + w^F_{dk} (1 - [2 w^I - 1] \cos\dmd t) \right]
      \right) \nonumber \\
 & \! + & \!\!\frac{f_s}{2} \frac{e^{-t/\tau_s}}{\tau_s}
      \left(\sum_{k=1}^{5}~g_{sk}
      \left[(1-w^F_{sk}) (1 + [2 w^I - 1] \cos\dms t)
           + w^F_{sk} (1 - [2 w^I - 1] \cos\dms t) \right]
      \right) \nonumber \\
 & \! + & \!\! f_{baryon} \frac{e^{-t/\tau_{baryon}}}{\tau_{baryon}}
      \left[w^I (1 - w^F_{baryon}) + (1 - w^I) w^F_{baryon} \right]
   \nonumber \\
 & \! + & \!\!\frac{f_{udsc}}{2}~F_{udsc}(t), \label{Pmixdipole}
\end{eqnarray}
where $f_j$ represents the fraction of each $b$-hadron type and background
($j=u,d,s$, $baryon$, and $udsc$ correspond to $\bu$, $\bd$, $\bs$, $b$-baryon,
and $udsc$ background),
$F_{udsc}(t)$ is a function describing the proper time distribution
of the $udsc$ background (a sum of two exponentials is used),
$\tau_j$ is the lifetime for $b$ hadrons of type $j$,
$w^I$ is the initial state mistag probability,
$w^F_u$ and $w^F_{baryon}$ are the final state mistag probabilities
for $\bu$ and $b$ baryons, whereas
$w^F_{dk}$ and $w^F_{sk}$ are the final state mistag probabilities
for $\bd$ and $\bs$, with the index $k= 1, ..., 5$
representing the five different decay final states:
$D^0 X$, $D^+ X$, $D_s X$, charmed hadron $X$, and
$D \overline{D} X$,
$g_{dk}$ and $g_{sk}$ are the fractions of $\bd$ and $\bs$ decays into
each of the above final states.
A similar expression for the probability ${\cal P}_{unmixed}$ to observe
a decay tagged as unmixed is obtained by replacing the initial state
mistag rate $w^I$ by $(1 - w^I)$.

  Several of the quantities in Eq.~(\ref{Pmixdipole}) are determined on
an event by event basis. The initial state mistag probability $w^I$ depends
on $\cos\theta$ of the thrust axis, the electron beam polarization, as well
as several quantities from the opposite hemisphere: jet charge, vertex charge,
kaon charge, lepton charge and dipole charge.
The final state mistag probabilities $w^F_{jk}$ depend on the reconstructed
decay length to take into account the degradation of the charge dipole tag
close to the IP. This effect is fairly weak for $\bd$, $\bs$ and
$b$ baryon decays (see, for example, Fig.~\ref{fig_dipo_tagpur}) but
it is significant for $\bu$ decays.
Finally, the decay final state fractions $g_{dk}$ and $g_{sk}$ are
parametrized as a function of the overall vertex mass $M$. For example,
the fraction of $\bs$ decays into $D \overline{D} X$ final states decreases
from 0.28 at $M = 2$ GeV/c$^2$ to 0.12 at $M = 5$ GeV/c$^2$.

  As described in Sec.~\ref{sec_lepd_likelihood}, the functions
${\cal P}_{mixed}$ and ${\cal P}_{unmixed}$ are convoluted with a proper time
resolution function ${\cal R}(T,t)$, see Eq.~(\ref{Resol})
and a time-dependent efficiency function $\varepsilon_j(t)$.
Separate efficiency functions are extracted for each $b$ hadron type using
the simulation.
The relative boost resolution $\sigma_{\gamma\beta}/{\gamma\beta}$
is parametrized with the sum of two Gaussians using the MC simulation.
Considering all selected $\bs$ decays,
the widths of the two Gaussians are
$\sigma_{B1} = 0.07$ and $\sigma_{B2} = 0.21$, where the first Gaussian
represents 60\% of the decays.
However, the analysis takes into account the strong dependence of the
resolution as a function of total charged track energy in each decay.
This is done separately for each of the four different $b$ hadron types.
Offsets in the boost reconstruction, especially
for decays with low reconstructed boost, have been corrected for as well.

The decay length resolution is also parametrized by the sum of two
Gaussians.
For decays with more than 1 track in either the $B$ or $D$ vertex,
the resolution $\sigma_L$ is estimated from the 
$B$ vertex fit and IP position measurement errors,
combined to yield an uncertainty $\sigma_{meas}$ along the flight direction.
The decay length resolution is then obtained by appropriately scaling
this quantity to determine a 60\% core resolution
$\sigma_{L1} = s_1 \times \sigma_{meas}$ and
a 40 \% tail resolution $\sigma_{L2} = s_2 \times \sigma_{meas}$.
For correctly tagged $\bs$ decays we find $s_1= 0.92$ and $s_2 = 2.30$.
For decays with 1 track in each of the $B$ and $D$ vertices,
the resolution is extracted from the overall decay length residual
distributions in the simulation.
The difference in decay length resolution between right and wrong
charge dipole tags motivates treating those separately in the likelihood
function, see Eq.~(\ref{Pmixdipole}).
(This is similar to differences in resolution between
$(b \to l)$ and $(b \to c \to l)$ in the lepton+D analysis.)
For example, the average decay length resolution for all $\bs$ decays
with right (wrong) charge dipole tag can be parameterized
by the sum of two Gaussians of widths
$\sigma_{L1} = 76\:\mu$m ($112\:\mu$m) and
$\sigma_{L2} = 311\:\mu$m ($450\:\mu$m), where the first Gaussian
represents 60\% of the decays.
In the case of $\bd$ and $\bs$ decays, the right- and wrong-tag
decay length resolutions are
estimated for each of the five decay final states.
Decays reconstructed within $200\:\mu$m of the IP have worse decay length
resolution and suffer from asymmetric tails (this is most likely due to
the addition of a primary track in the secondary vertex).
These effects have been taken into account in the analysis.
Finally, offsets in the reconstructed decay length are corrected
separately for decays involving one or two charm particles, as the effect
is small for the former but not negligible for the latter.

\subsection{Oscillation Analysis}

\begin{figure}[t]
  \hspace*{6mm}
  \centering
  \epsfxsize=11cm
  \epsfbox{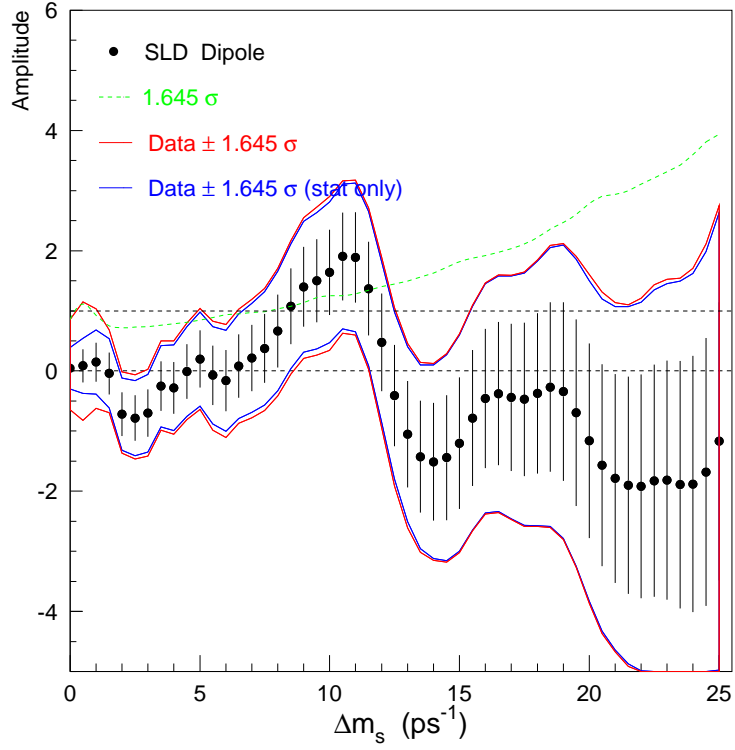}
  \caption{\it \label{fig_dipo_afit}
  \baselineskip=12pt
  Measured amplitude as a function of $\dms$ in the charge dipole analysis.}
  \baselineskip=18pt
\end{figure}

An amplitude fit is performed, as described in Sec.~\ref{sec_lepd_afit} and
the result is displayed in Fig.~\ref{fig_dipo_afit}.
Systematic uncertainties are estimated as for the lepton+D analysis
except for those affecting the final state tag.
Here, uncertainties in the charge dipole correct tag probability
modeling are obtained by varying the scale factor applied to the probability
derived from the MC simulation according to $S_D = 0.95 \pm 0.025$.
In addition, the systematic uncertainty also includes the effect
due to a further reduction of the scale factor by 5\%
for decays within $L < 0.5$ mm of the IP.
Dominant uncertainties are due to the $\bs$ production fraction
in $\Zbb$ events, the boost and decay length resolutions,
and the overall uncertainty in the final state
tag purity, see Table~\ref{tbl_dipo_syst}.
\begin{table}[t]
\caption{Measured values of the oscillation amplitude $A$ with a breakdown
   of the main systematic uncertainties for several $\dms$ values
   in the charge dipole analysis.}
\begin{center}
\begin{tabular}{lcccc}
 $\dms$                  & ~~5 ps$^{-1}$  & ~~10 ps$^{-1}$  & ~~15 ps$^{-1}$ 
                         & ~~20 ps$^{-1}$ \\
 \hline
 \vspace{0.1cm}
 Measured amplitude $A$  &       ~0.197   &       ~1.639   &      $-1.201$ 
                         &      $-1.162$ \\
 \vspace{0.1cm}
 $\sigma_A^{stat}$       &   $\pm 0.476$  &   $\pm 0.711$  &   $\pm 1.092$
                         &   $\pm 1.620$  \\
 \vspace{0.1cm}
 $\sigma_A^{syst}$       &
         $^{+0.197}_{-0.178}$ & $^{+0.268}_{-0.351}$ & $^{+0.159}_{-0.187}$
       & $^{+0.489}_{-0.189}$ \\
 \hline
 \vspace{0.2cm}
 $f_s = {\cal B}(\bar{b} \to \bs)$ &
         $^{-0.134}_{+0.172}$ & $^{-0.116}_{+0.152}$ & $^{-0.105}_{+0.130}$
       & $^{-0.111}_{+0.116}$ \\
 \vspace{0.1cm}
 $f_\Lambda = {\cal B}(b \to b{\rm -baryon})$ &
         $^{+0.015}_{-0.013}$ & $^{+0.020}_{-0.019}$ & $^{+0.020}_{-0.020}$
       & $^{+0.008}_{-0.019}$ \\
 \vspace{0.1cm}
 $udsc$ fraction &
         $^{-0.004}_{+0.004}$ & $^{-0.019}_{+0.025}$ & $^{-0.025}_{+0.021}$ 
       & $^{+0.001}_{+0.004}$ \\
 \vspace{0.1cm}
 decay length resolution &
         $^{+0.017}_{-0.020}$ & $^{+0.048}_{-0.035}$ & $^{-0.026}_{-0.002}$ 
       & $^{+0.080}_{-0.047}$ \\
 \vspace{0.1cm}
 boost resolution &
         $^{+0.049}_{-0.044}$ & $^{+0.209}_{-0.227}$ & $^{+0.075}_{-0.096}$ 
       & $^{+0.450}_{-0.084}$ \\
 \vspace{0.1cm}
 $\bs$ lifetime &
         $^{+0.039}_{-0.038}$ & $^{+0.014}_{-0.017}$ & $^{+0.030}_{-0.029}$ 
       & $^{+0.028}_{-0.045}$ \\
 \vspace{0.1cm}
 $\dmd$ &
         $^{+0.002}_{-0.001}$ & $^{+0.001}_{-0.001}$ & $^{+0.000}_{-0.002}$ 
       & $^{-0.011}_{+0.001}$ \\
 \vspace{0.1cm}
 initial state tag &
         $^{+0.001}_{+0.000}$ & $^{+0.016}_{-0.014}$ & $^{+0.004}_{-0.002}$ 
       & $^{-0.106}_{+0.099}$ \\
 \vspace{0.1cm}
 final state tag &
         $^{+0.069}_{-0.099}$ & $^{+0.036}_{-0.236}$ & $^{+0.027}_{-0.110}$ 
       & $^{+0.074}_{-0.008}$ \\
 \hline
\end{tabular}
\label{tbl_dipo_syst}
\end{center}
\end{table}

\pagebreak

\section{Combination of the Analyses}
\begin{figure}[p]
  \hspace*{6mm}
  \centering
  \epsfxsize=14cm
  \epsfbox{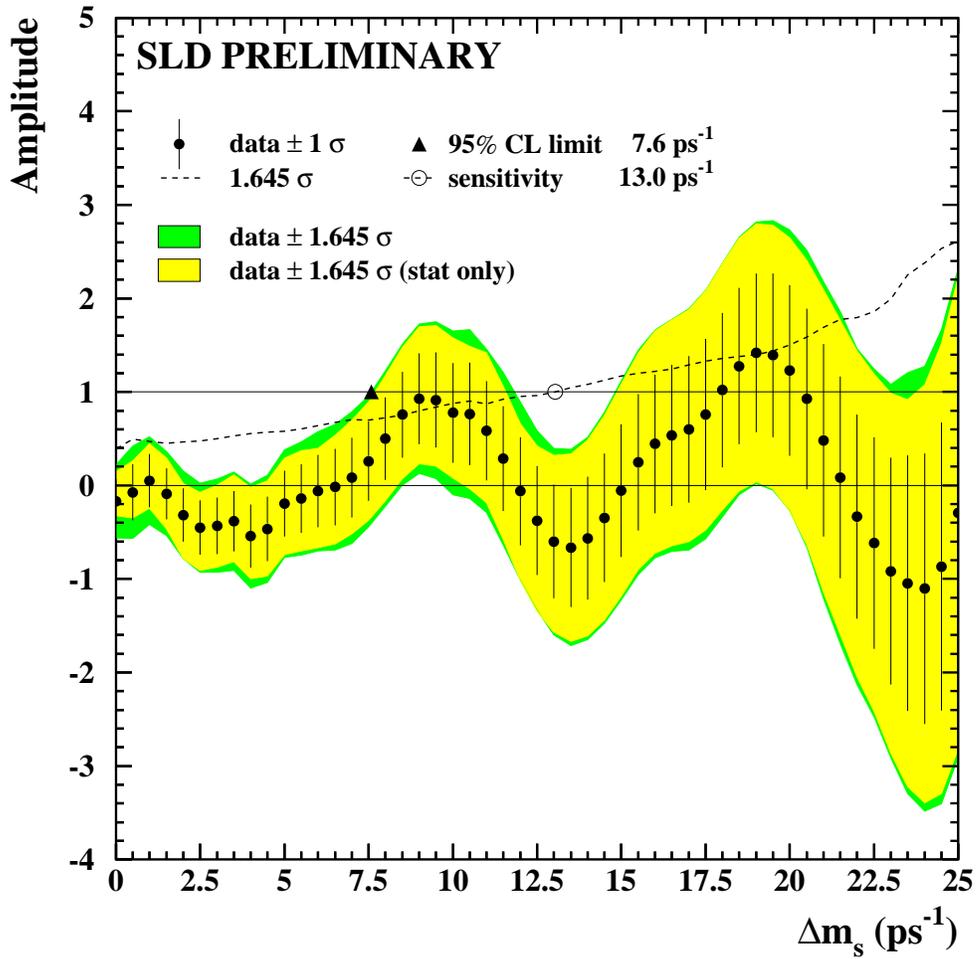}
  \caption{\it \label{fig_comb_afit}
  \baselineskip=12pt
  Measured amplitude as a function of $\dms$ for the lepton+D,
  D$_{\rm s}$+tracks, and charge dipole analyses combined.}
  \baselineskip=18pt
\end{figure}

The D$_{\rm s}$+tracks, lepton+D and charge dipole analyses are
combined taking into account correlated systematic errors.
Events shared by two or more analyses are assigned to the analysis
with the best sensitivity such as to produce statistically independent
analyses.
Figure~\ref{fig_comb_afit} shows the measured amplitude as a function
of $\dms$ for the combination.
As noted earlier, the measured values are consistent with $A = 0$
for the whole range of $\dms$ up to 25 ps$^{-1}$
and no evidence is found for a preferred value of the mixing frequency.
The following ranges of $\bsmix$ oscillation frequencies
are excluded at 95\% C.L.:
$\Delta m_s < 7.6$ ps$^{-1}$ and
$11.8 < \Delta m_s < 14.8$ ps$^{-1}$, i.e.,
the condition $A + 1.645\:\sigma_A < 1$ is satisfied for those values.
The combined sensitivity to set a 95\% C.L. lower limit is
found to be at a $\dms$ value of 13.0 ps$^{-1}$.
These results are preliminary.

\section*{Acknowledgments}

        We thank the personnel of the SLAC accelerator department and
the technical staffs of our collaborating institutions for their outstanding
efforts.

\pagebreak

%
%
%
\section*{$^{**}$ List of Authors}

\begin{center}
\def\iAOMORI{$^{(1)}$}
\def\iBRI{$^{(2)}$}
\def\iBRUN{$^{(3)}$}
\def\iBU{$^{(4)}$}
\def\iCOLO{$^{(5)}$}
\def\iCSU{$^{(6)}$}
\def\iFERR{$^{(7)}$}
\def\iFRAS{$^{(8)}$}
\def\iJHU{$^{(9)}$}
\def\iLBL{$^{(10)}$}
\def\iMASS{$^{(11)}$}
\def\iMISSI{$^{(12)}$}
\def\iMIT{$^{(13)}$}
\def\iMOSCOW{$^{(14)}$}
\def\iNAGO{$^{(15)}$}
\def\iOREG{$^{(16)}$}
\def\iOXF{$^{(17)}$}
\def\iPERU{$^{(18)}$}
\def\iRAL{$^{(19)}$}
\def\iRUTG{$^{(20)}$}
\def\iSLAC{$^{(21)}$}
\def\iSOONG{$^{(22)}$}
\def\iTENN{$^{(23)}$}
\def\iTOHO{$^{(24)}$}
\def\iUCSB{$^{(25)}$}
\def\iUCSC{$^{(26)}$}
\def\iVAND{$^{(27)}$}
\def\iWASH{$^{(28)}$}
\def\iWISC{$^{(29)}$}
\def\iYALE{$^{(30)}$}

  \baselineskip=.75\baselineskip  
\mbox{Kenji Abe\unskip,\iNAGO}
\mbox{Koya Abe\unskip,\iTOHO}
\mbox{T. Abe\unskip,\iSLAC}
\mbox{I. Adam\unskip,\iSLAC}
\mbox{H. Akimoto\unskip,\iSLAC}
\mbox{D. Aston\unskip,\iSLAC}
\mbox{K.G. Baird\unskip,\iMASS}
\mbox{C. Baltay\unskip,\iYALE}
\mbox{H.R. Band\unskip,\iWISC}
\mbox{T.L. Barklow\unskip,\iSLAC}
\mbox{J.M. Bauer\unskip,\iMISSI}
\mbox{G. Bellodi\unskip,\iOXF}
\mbox{R. Berger\unskip,\iSLAC}
\mbox{G. Blaylock\unskip,\iMASS}
\mbox{J.R. Bogart\unskip,\iSLAC}
\mbox{G.R. Bower\unskip,\iSLAC}
\mbox{J.E. Brau\unskip,\iOREG}
\mbox{M. Breidenbach\unskip,\iSLAC}
\mbox{W.M. Bugg\unskip,\iTENN}
\mbox{D. Burke\unskip,\iSLAC}
\mbox{T.H. Burnett\unskip,\iWASH}
\mbox{P.N. Burrows\unskip,\iOXF}
\mbox{A. Calcaterra\unskip,\iFRAS}
\mbox{R. Cassell\unskip,\iSLAC}
\mbox{A. Chou\unskip,\iSLAC}
\mbox{H.O. Cohn\unskip,\iTENN}
\mbox{J.A. Coller\unskip,\iBU}
\mbox{M.R. Convery\unskip,\iSLAC}
\mbox{V. Cook\unskip,\iWASH}
\mbox{R.F. Cowan\unskip,\iMIT}
\mbox{G. Crawford\unskip,\iSLAC}
\mbox{C.J.S. Damerell\unskip,\iRAL}
\mbox{M. Daoudi\unskip,\iSLAC}
\mbox{S. Dasu\unskip,\iWISC}
\mbox{N. de Groot\unskip,\iBRI}
\mbox{R. de Sangro\unskip,\iFRAS}
\mbox{D.N. Dong\unskip,\iMIT}
\mbox{M. Doser\unskip,\iSLAC}
\mbox{R. Dubois\unskip,\iSLAC}
\mbox{I. Erofeeva\unskip,\iMOSCOW}
\mbox{V. Eschenburg\unskip,\iMISSI}
\mbox{S. Fahey\unskip,\iCOLO}
\mbox{D. Falciai\unskip,\iFRAS}
\mbox{J.P. Fernandez\unskip,\iUCSC}
\mbox{K. Flood\unskip,\iMASS}
\mbox{R. Frey\unskip,\iOREG}
\mbox{E.L. Hart\unskip,\iTENN}
\mbox{K. Hasuko\unskip,\iTOHO}
\mbox{S.S. Hertzbach\unskip,\iMASS}
\mbox{M.E. Huffer\unskip,\iSLAC}
\mbox{X. Huynh\unskip,\iSLAC}
\mbox{M. Iwasaki\unskip,\iOREG}
\mbox{D.J. Jackson\unskip,\iRAL}
\mbox{P. Jacques\unskip,\iRUTG}
\mbox{J.A. Jaros\unskip,\iSLAC}
\mbox{Z.Y. Jiang\unskip,\iSLAC}
\mbox{A.S. Johnson\unskip,\iSLAC}
\mbox{J.R. Johnson\unskip,\iWISC}
\mbox{R. Kajikawa\unskip,\iNAGO}
\mbox{M. Kalelkar\unskip,\iRUTG}
\mbox{H.J. Kang\unskip,\iRUTG}
\mbox{R.R. Kofler\unskip,\iMASS}
\mbox{R.S. Kroeger\unskip,\iMISSI}
\mbox{M. Langston\unskip,\iOREG}
\mbox{D.W.G. Leith\unskip,\iSLAC}
\mbox{V. Lia\unskip,\iMIT}
\mbox{C. Lin\unskip,\iMASS}
\mbox{G. Mancinelli\unskip,\iRUTG}
\mbox{S. Manly\unskip,\iYALE}
\mbox{G. Mantovani\unskip,\iPERU}
\mbox{T.W. Markiewicz\unskip,\iSLAC}
\mbox{T. Maruyama\unskip,\iSLAC}
\mbox{A.K. McKemey\unskip,\iBRUN}
\mbox{R. Messner\unskip,\iSLAC}
\mbox{K.C. Moffeit\unskip,\iSLAC}
\mbox{T.B. Moore\unskip,\iYALE}
\mbox{M. Morii\unskip,\iSLAC}
\mbox{D. Muller\unskip,\iSLAC}
\mbox{V. Murzin\unskip,\iMOSCOW}
\mbox{S. Narita\unskip,\iTOHO}
\mbox{U. Nauenberg\unskip,\iCOLO}
\mbox{H. Neal\unskip,\iYALE}
\mbox{G. Nesom\unskip,\iOXF}
\mbox{N. Oishi\unskip,\iNAGO}
\mbox{D. Onoprienko\unskip,\iTENN}
\mbox{L.S. Osborne\unskip,\iMIT}
\mbox{R.S. Panvini\unskip,\iVAND}
\mbox{C.H. Park\unskip,\iSOONG}
\mbox{I. Peruzzi\unskip,\iFRAS}
\mbox{M. Piccolo\unskip,\iFRAS}
\mbox{L. Piemontese\unskip,\iFERR}
\mbox{R.J. Plano\unskip,\iRUTG}
\mbox{R. Prepost\unskip,\iWISC}
\mbox{C.Y. Prescott\unskip,\iSLAC}
\mbox{B.N. Ratcliff\unskip,\iSLAC}
\mbox{J. Reidy\unskip,\iMISSI}
\mbox{P.L. Reinertsen\unskip,\iUCSC}
\mbox{L.S. Rochester\unskip,\iSLAC}
\mbox{P.C. Rowson\unskip,\iSLAC}
\mbox{J.J. Russell\unskip,\iSLAC}
\mbox{O.H. Saxton\unskip,\iSLAC}
\mbox{T. Schalk\unskip,\iUCSC}
\mbox{B.A. Schumm\unskip,\iUCSC}
\mbox{J. Schwiening\unskip,\iSLAC}
\mbox{V.V. Serbo\unskip,\iSLAC}
\mbox{G. Shapiro\unskip,\iLBL}
\mbox{N.B. Sinev\unskip,\iOREG}
\mbox{J.A. Snyder\unskip,\iYALE}
\mbox{H. Staengle\unskip,\iCSU}
\mbox{A. Stahl\unskip,\iSLAC}
\mbox{P. Stamer\unskip,\iRUTG}
\mbox{H. Steiner\unskip,\iLBL}
\mbox{D. Su\unskip,\iSLAC}
\mbox{F. Suekane\unskip,\iTOHO}
\mbox{A. Sugiyama\unskip,\iNAGO}
\mbox{S. Suzuki\unskip,\iNAGO}
\mbox{M. Swartz\unskip,\iJHU}
\mbox{F.E. Taylor\unskip,\iMIT}
\mbox{J. Thom\unskip,\iSLAC}
\mbox{E. Torrence\unskip,\iMIT}
\mbox{T. Usher\unskip,\iSLAC}
\mbox{J. Va'vra\unskip,\iSLAC}
\mbox{R. Verdier\unskip,\iMIT}
\mbox{D.L. Wagner\unskip,\iCOLO}
\mbox{A.P. Waite\unskip,\iSLAC}
\mbox{S. Walston\unskip,\iOREG}
\mbox{A.W. Weidemann\unskip,\iTENN}
\mbox{E.R. Weiss\unskip,\iWASH}
\mbox{J.S. Whitaker\unskip,\iBU}
\mbox{S.H. Williams\unskip,\iSLAC}
\mbox{S. Willocq\unskip,\iMASS}
\mbox{R.J. Wilson\unskip,\iCSU}
\mbox{W.J. Wisniewski\unskip,\iSLAC}
\mbox{J.L. Wittlin\unskip,\iMASS}
\mbox{M. Woods\unskip,\iSLAC}
\mbox{T.R. Wright\unskip,\iWISC}
\mbox{R.K. Yamamoto\unskip,\iMIT}
\mbox{J. Yashima\unskip,\iTOHO}
\mbox{S.J. Yellin\unskip,\iUCSB}
\mbox{C.C. Young\unskip,\iSLAC}
\mbox{H. Yuta\unskip.\iAOMORI}

\it
  \vskip \baselineskip                   
  \centerline{(The SLD Collaboration)}   
  \vskip \baselineskip        
  \baselineskip=.75\baselineskip   
\iAOMORI
  Aomori University, Aomori , 030 Japan, \break
\iBRI
  University of Bristol, Bristol, United Kingdom, \break
\iBRUN
  Brunel University, Uxbridge, Middlesex, UB8 3PH United Kingdom, \break
\iBU
  Boston University, Boston, Massachusetts 02215, \break
\iCOLO
  University of Colorado, Boulder, Colorado 80309, \break
\iCSU
  Colorado State University, Ft. Collins, Colorado 80523, \break
\iFERR
  INFN Sezione di Ferrara and Universita di Ferrara, I-44100 Ferrara, Italy, \break
\iFRAS
  INFN Lab. Nazionali di Frascati, I-00044 Frascati, Italy, \break
\iJHU
  Johns Hopkins University,  Baltimore, Maryland 21218-2686, \break
\iLBL
  Lawrence Berkeley Laboratory, University of California, Berkeley, California 94720, \break
\iMASS
  University of Massachusetts, Amherst, Massachusetts 01003, \break
\iMISSI
  University of Mississippi, University, Mississippi 38677, \break
\iMIT
  Massachusetts Institute of Technology, Cambridge, Massachusetts 02139, \break
\iMOSCOW
  Institute of Nuclear Physics, Moscow State University, 119899, Moscow Russia, \break
\iNAGO
  Nagoya University, Chikusa-ku, Nagoya, 464 Japan, \break
\iOREG
  University of Oregon, Eugene, Oregon 97403, \break
\iOXF
  Oxford University, Oxford, OX1 3RH, United Kingdom, \break
\iPERU
  INFN Sezione di Perugia and Universita di Perugia, I-06100 Perugia, Italy, \break
\iRAL
  Rutherford Appleton Laboratory, Chilton, Didcot, Oxon OX11 0QX United Kingdom, \break
\iRUTG
  Rutgers University, Piscataway, New Jersey 08855, \break
\iSLAC
  Stanford Linear Accelerator Center, Stanford University, Stanford, California 94309, \break
\iSOONG
  Soongsil University, Seoul, Korea 156-743, \break
\iTENN
  University of Tennessee, Knoxville, Tennessee 37996, \break
\iTOHO
  Tohoku University, Sendai 980, Japan, \break
\iUCSB
  University of California at Santa Barbara, Santa Barbara, California 93106, \break
\iUCSC
  University of California at Santa Cruz, Santa Cruz, California 95064, \break
\iVAND
  Vanderbilt University, Nashville,Tennessee 37235, \break
\iWASH
  University of Washington, Seattle, Washington 98105, \break
\iWISC
  University of Wisconsin, Madison,Wisconsin 53706, \break
\iYALE
  Yale University, New Haven, Connecticut 06511. \break

\rm
%

\end{center}


\begin{thebibliography}{99}

\bibitem{Stocchi} See the following reviews: F. Caravaglios, F. Parodi,
 P.~Roudeau, and A.~Stocchi,
 {\em Determination of the CKM unitarity triangle parameters by end 1999},
 {\tt hep-ph/0002171};
 S.~Mele, Phys. Rev. {\bf D59}, 113011 (1999).

\bibitem{Hashimoto}
 S. Hashimoto, {\em B decays on the lattice},
 {\tt hep-lat/9909136}, Nucl.Phys.Proc.Suppl. 83, 3 (2000).

 \bibitem{Dstracks} K.~Abe {\it et al.},
 {\it Time Dependent $B_s^0 - \overline{B_s^0}$ Oscillations
 Using Exclusively Reconstructed $D_s^+$ Decays at SLD},
 SLAC-PUB-8598, August 2000.

 \bibitem{rbrb}  K.~Abe {\it et~al.}, Phys. Rev. {\bf D53}, 1023 (1996).

 \bibitem{vxd3} K.~Abe {\it et~al.},
     Nucl. Inst. and Meth. {\bf A400}, 287 (1997).

 \bibitem{jetset} T. Sj\"{o}strand, Comp. Phys. Comm.
 {\bf 82}, 74 (1994).

 \bibitem{CLEO-QQ} CLEO $B$ decay model provided by P.~Kim and the
                   CLEO Collaboration.

 \bibitem{argcl}  B.~Barish {\it et~al.},
                  Phys. Rev. Lett. {\bf 76}, 1570 (1996);  
                  H.~Albrecht {\it et~al.},
                  Z. Phys. {\bf C58}, 191 (1993);
                  H.~Albrecht {\it et~al.},
                  Z. Phys. {\bf C62}, 371 (1994);
                  P.~Avery {\it et al.}, CLEO CONF 96-28, July 1996;
                  L.~Gibbons {\it et al.}, Phys. Rev. {\bf D56}, 3783 (1997);
                  T.E.~Coan {\it et al.}, CLNS 97/1516;
                  CLEO Collab., CLEO CONF 97-27, Aug. 1997;
                  M.~Zoeller, Ph.D. Thesis, SUNY Albany, 1994;
                  X.~Fu {\it et al.}, Phys. Rev. Lett. {\bf 79}, 3125 (1997);
                  D.~Gibaut {\it et al.}, Phys. Rev. {\bf D53}, 4734 (1996).

 \bibitem{ISGW} N.~Isgur, D.~Scora, B.~Grinstein, and M.B.~Wise,
                Phys. Rev. {\bf D39}, 799 (1989).

 \bibitem{PDG96} Particle Data Group, Phys. Rev. {\bf D54}, Part I (1996).

 \bibitem{LEPHF99} D. Abbaneo {\it et al.} (LEP Heavy Flavor Steering Group),
  {\sl Combined Results on b Hadron Production Rates, Lifetimes, Oscillations
  and Semileptonic Decays}, CERN-EP-2000-096, SLAC-PUB-8492, March 2000.
  

 \bibitem{Peterson} C. Peterson {\it et al.}, Phys. Rev. {\bf D27}, 105 (1983).

 \bibitem{geant} R. Brun {\it et al.}, Report No. CERN-DD/EE/84-1, 1989.

 \bibitem{zvtop} D. J. Jackson,
     Nucl. Inst. and Meth. {\bf A388}, 247 (1997).

\bibitem{blife} K.~Abe {\it et al.} (SLD Collaboration),
 {\em Measurement of the $B^+$ and $B^0$ Lifetimes using Topological Vertexing
 at SLD},
 SLAC-PUB-8206, July 1999,
 contributed paper \# 477 to EPS-HEP99.

 \bibitem{rbprl}  K.~Abe {\it et~al.},
     Phys. Rev. Lett. {\bf 80}, 660 (1998).

\bibitem{bfrag}  K.~Abe {\it et~al.},
     Phys. Rev. {\bf D56}, 5310 (1997).

\bibitem{Moser} H.-G.~Moser and A.~Roussarie,
                Nucl.~Inst. and Meth. {\bf A384}, 491 (1997).

\bibitem{golutvin}
 See A. Golutvin, {\it Heavy Flavour Physics}, summary talk presented
at ICHEP 2000.
                          
\end{thebibliography}
\enddocument